\documentclass[11pt,a4paper]{article}
\usepackage{geometry}
\usepackage{graphicx}
\usepackage{bm,array}
\usepackage{comment}
\usepackage{caption}
\usepackage{subcaption}
\usepackage{tablefootnote}
 \usepackage{makecell}
\usepackage{booktabs}
 \usepackage{multirow}
 \usepackage{float}
 \usepackage{framed}
 \usepackage[ruled,vlined,linesnumbered]{algorithm2e}
\usepackage{algorithmic}

\usepackage[normalem]{ulem}

\usepackage[pdfpagemode={UseOutlines},bookmarks=true,bookmarksopen=true,
bookmarksopenlevel=0,bookmarksnumbered=true,hypertexnames=false,
colorlinks,linkcolor={blue},citecolor={blue},urlcolor={red},
pdfstartview={FitV},unicode,breaklinks=true]{hyperref}
\hypersetup{urlcolor=blue, colorlinks=true}

\usepackage{setspace}
\usepackage{vmargin}
\setmarginsrb{ 1.0in}  
{ 0.6in}  
{ 1.0in}  
{ 0.8in}  
{  20pt}  
{0.25in}  
{9pt}  
{ 0.3in}  

\usepackage{authblk}

\usepackage{amsmath}
\usepackage{amsfonts}
\usepackage{amssymb}
\usepackage[round, sort,comma,authoryear]{natbib}

\usepackage{mdframed}
\usepackage{tikz}
\usetikzlibrary{arrows.meta,positioning,shapes.geometric}

\definecolor{lightgray}{rgb}{0.9,0.9,0.9}

\newtheorem{theorem}{Theorem}

\newtheorem{lemma}{Lemma}

\newtheorem{proposition}{Proposition}

\newenvironment{proof}[1][Proof]{\noindent\textbf{#1.} }{\ \rule{0.5em}{0.5em}}

\usepackage{multirow}
\usepackage{hhline}
\usepackage{subcaption}
\usepackage{footnote}
\usepackage[ruled,vlined]{algorithm2e}
\usepackage{algorithmic}

\usepackage{empheq}

\newcommand{\MNL}{\textsc{MNL}}
\newcommand{\transpose}{{\mbox{\tiny T}}}

\newcommand{\cS}{{\mathcal{S}}}

\newcommand{\cL}{{\mathcal{L}}}

\newcommand{\cK}{{\mathcal{K}}}

\newcommand{\cN}{{\mathcal{N}}}

\newcommand{\cI}{{\mathcal{I}}}

\newcommand{\by}{\textbf{y}}

\newcommand{\ba}{\textbf{a}}

\newcommand{\bz}{\textbf{z}}

\newcommand{\bt}{\textbf{t}}
 
\newcommand{\bbt}{\pmb{\beta}}
\newcommand{\bld}{{\pmb{\lambda}}}

\newcommand{\blambda}{\pmb{\lambda}}
\newcommand{\beps}{\pmb{\epsilon}}
\newcommand{\MLEvD}{\texttt{[MLE via Discretization]}}
\newcommand{\argmax}{\text{argmax}}

\newcommand{\bbR}{\mathbb{R}}

\newcommand{\TNL}{\textsc{\tiny TNL}}
\newcommand{\NL}{\textsc{\tiny NL}}

\usepackage{mathtools}

\usepackage{mathtools}

\newif\ifnotes\notestrue
\def\mtien#1{{\color{magenta}{#1}}}

\def\htien#1{}

\usepackage{xcolor}

\begin{document}

\newcolumntype{C}{>{\centering\arraybackslash}p{4em}}

\spacing{1.5}

\title{\textbf{On the Estimation of Multinomial Logit and Nested Logit Models: A Conic Optimization Approach }}
\author[1]{Hoang Giang Pham}
\author[1,*]{Tien Mai}
\author[2]{Minh Ha Hoang}
\affil[1]{\it\small
School of Computing and Information Systems, Singapore Management University}
\affil[2]{\it\small
SLSCM and CADA, Faculty of Data Science and Artificial Intelligence, College of Technology, National Economics University, Hanoi, Vietnam}
\affil[*]{\it\small
Corresponding author, atmai@smu.edu.sg}

\date{}
\maketitle

\begin{abstract}
In this paper, we revisit parameter estimation for multinomial logit (MNL), nested logit (NL), and tree–nested logit (TNL) models through the framework of \textit{convex conic optimization}. Traditional approaches typically solve the maximum-likelihood estimation (MLE) problem using gradient-based methods, which are sensitive to step-size selection and initialization, and may therefore suffer from slow or unstable convergence. In contrast, we propose a novel estimation strategy that reformulates these models as conic optimization problems, enabling more robust and reliable estimation procedures.
Specifically, we show that the MLE for MNL admits an equivalent \emph{exponential cone program} (ECP). For NL and TNL, we prove that, when the dissimilarity (scale) parameters  are fixed, the estimation problem is convex and likewise reducible to an ECP. Leveraging these results, we design a two–stage procedure: an outer loop that updates the scale parameters and an inner loop that solves the ECP to update the utility coefficients. The inner problems are handled by interior-point methods with iteration counts that grow only logarithmically in the target accuracy, as implemented in off-the-shelf solvers (e.g., MOSEK). Extensive experiments across estimation instances of varying size show that our conic approach attains better MLE solutions, greater robustness to initialization, and substantial  speedups compared to standard gradient-based MLE, particularly on large-scale instances with high-dimensional specifications and large choice sets. Our findings establish exponential-cone programming as a practical and scalable alternative for estimating a broad class of discrete-choice models.
\end{abstract}



\section{Introduction}

Discrete choice models are fundamental tools for analyzing individual decision-making in domains such as transportation, marketing, industrial organization, and operations research. Since the seminal contributions of \citet{McFadden1974} and \citet{BenAkivaLerman1985}, the multinomial logit (MNL), nested logit (NL), and tree--nested logit (TNL) models have become standard due to their behavioral interpretability, closed-form choice probabilities, and analytical tractability \citep{Train2009}. Estimation is typically carried out via maximum likelihood, solved with gradient-based routines such as Newton, quasi-Newton, or sequential quadratic programming methods \citep{BerndtEtAl1974,NocedalWright2006}. While these methods are effective in small- to medium-scale applications, they can be sensitive to step-size selection and initialization, and their numerical performance often deteriorates in large-scale settings with high-dimensional covariates, large choice sets, or complex nesting structures \citep{BorschSupanHajivassiliou1993,Train2009,HessDaly2014}.

This paper revisits estimation of MNL, NL, and TNL through the lens of \textit{convex conic optimization}. We show that the MLE for the MNL model admits an exact reformulation as an \emph{exponential-cone program} (ECP) \citep{BoydVandenberghe2004,Nesterov1994interior,MOSEKcookbook}. For NL and TNL, when the dissimilarity (scale) parameters are fixed, the MLE objective is convex and can likewise be written as an ECP. Building on these observations, we propose a two-stage procedure that alternates between (i) an outer update of the nesting scale parameters and (ii) an inner ECP that updates the utility coefficients.

Casting estimation as an ECP brings algorithmic and practical benefits. First, it enables the use of primal–dual interior-point methods with iteration counts that grow only logarithmically with the desired accuracy \citep{Nesterov1994interior,BoydVandenberghe2004}. Second, off-the-shelf solvers implementing these methods \citep{MOSEKcookbook} can exploit the block structure induced by observations and alternatives, delivering robust performance without delicate hand-tuning of step sizes or line-search heuristics. Third, the conic viewpoint yields unified formulations across MNL, NL, and TNL, simplifying implementation and facilitating large-scale estimation.

We evaluate the proposed approach on synthetic datasets spanning a wide range of dimensions, numbers of nests, and choice-set sizes. Across all regimes, the ECP reformulations consistently deliver better MLE solutions and substantially lower wall-clock times than strong gradient-based baselines, with the largest gains appearing on medium- and large-scale instances. The results underscore that exponential-cone programming is a practical, scalable, and numerically stable alternative to standard gradient-based MLE for a broad class of discrete-choice models.

\paragraph{Contributions.} 
This paper makes the following key contributions. 
First, we provide an exact ECP reformulation of the maximum-likelihood estimator for the MNL model. 
Second, we develop convex ECP reformulations for NL and TNL when the dissimilarity parameters are fixed, and propose a two-stage procedure that alternates between updating the scale parameters and solving the convex inner problem for the utility coefficients. 
Third, we present an extensive empirical study showing that our approach delivers superior scalability, robustness, and solution quality compared to standard gradient-based estimators. \textit{To the best of our knowledge, this is the first work to formulate and solve the estimation 
of several widely used discrete choice models using exponential-cone programming, achieving 
state-of-the-art performance compared to standard gradient-based methods.
 Moreover, 
building on these findings, we are the first to establish polynomial-time guarantees for 
estimating the MNL model, as well as the NL and TNL models with fixed scale parameters.
}

\paragraph{Organization.} 
The remainder of the paper is organized as follows.  Section \ref{sec:review} presents a literature review.
Section~\ref{sec:MNL} introduces the MNL formulation and its ECP reformulation. 
Sections~\ref{sec:NL} and \ref{sec:TNL} extend the approach to NL and TNL under fixed scale parameters and describes the two-stage estimation scheme. 
Section~\ref{sec:results} reports numerical results, and Section~\ref{sec:concl} concludes with a discussion of implications and future research directions.
The appendix contains proofs omitted from the main paper, along with detailed numerical analyses.

\paragraph{Notation:}
Boldface characters represent matrices (or vectors), and $a_i$ denotes the $i$-th element of vector $\ba$. We use $[m]$, for any $m\in \mathbb{N}$, to denote the set $\{1,\ldots,m\}$.

\section{Literature Review}
\label{sec:review}

Discrete-choice models have a long history as empirical workhorses in transportation, marketing, and operations. The MNL model occupies a central place because of its interpretability and analytical tractability under i.i.d.\ Type-I extreme value errors \citep{McFadden1974,BenAkivaLerman1985,Train2009}. The nested logit  and tree–nested logit models generalize the framework to accommodate richer substitution patterns when alternatives share unobserved components, using generating functions from the Generalized Extreme Value (GEV) family \citep{Williams1977,DalyZachary1978,Train2009}. Across these models, estimation is typically framed as maximum likelihood, with algorithms built around smooth optimization of the log-likelihood and its derivatives.

For the MNL model, the log-likelihood is strictly concave in the utility coefficients, which makes maximum-likelihood estimation well-posed and enables the use of Newton, Fisher scoring/BHHH, and quasi-Newton methods such as L-BFGS \citep{BerndtEtAl1974,LiuNocedal1989,Train2009}. Practical implementations emphasize efficient evaluation of utilities and choice probabilities across large numbers of observations and alternatives, as well as robust variance estimation \citep{Train2009}. When the universe of alternatives is very large, sampling-of-alternatives with correction terms becomes important to reduce computational burden while preserving consistency \citep[Ch.~3]{BenAkivaLerman1985}. Regularization has also become common, with $\ell_2$ penalties stabilizing estimates and $\ell_1$ penalties enabling high-dimensional specifications \citep[e.g.,][]{HastieTibshiraniFriedman2009}. Despite these advances, performance can deteriorate when covariates are poorly scaled, when choice sets are massive, or when gradients are ill-conditioned, making step-size selection, line searches, and preconditioning critical in practice \citep{NocedalWright2006}.

It is well known that the MNL model satisfies the independence from irrelevant alternatives (IIA) property, which states that the relative odds of choosing between two alternatives are unaffected by the presence or characteristics of other options. This assumption is often unrealistic in practice. The NL model relaxes IIA by allowing alternatives within a nest to share unobserved components \citep{Williams1977,DalyZachary1978}.
From an estimation perspective, the likelihood is concave in the utility coefficients conditional on the nest dissimilarity (scale) parameters, but the joint problem over both sets of parameters is not generally concave \citep{Train2009,DaganzoKusnic1993_convex_nested}. This has led to two common strategies: fixing the scale parameters based on prior evidence and estimating the utilities by standard MLE, or alternating between updates of the scales and conditional concave updates of the utilities \citep{Train2009}. To ensure identification, the NL model requires normalizing the overall scale and restricting the dissimilarity parameters to lie within valid bounds. In practice, estimation routines impose these conditions using box or nonlinear constraints \citep{Train2009,NoceWrig06}. The use of analytical “inclusive value” (logsum) formulas helps keep computations manageable, but the extra constraints and nonconvexity make estimation more challenging and less straightforward than in the MNL model \citep{BenAkivaLerman1985,Train2009}.

The TNL extends the NL model to multi-level correlation structures, attaching scale parameters to internal nodes along each branch of a choice hierarchy \citep[Ch.~3]{Train2009,Daly1987}. This added flexibility improves behavioral realism in settings with deep category structures, but it also complicates estimation: the likelihood remains convex in the utility coefficients for fixed scales, while the full problem (utilities plus all scale parameters) is generally nonconvex and subject to additional admissibility restrictions along each path of the tree \citep{Train2009,DaganzoKusnic1993_convex_nested}. As the depth and breadth of the nesting structure grow, conditioning worsens and gradient-based methods can become sensitive to initialization and step-size rules. In applied work, it is therefore common either to fix a subset of scales or to employ alternating or two-step procedures that separate scale updates from utility estimation to preserve conditional concavity \citep{Train2009,Daly1987}.

Beyond the MNL and NL models, two important generalizations are the cross-nested logit (CNL) \citep{Bier06} and the mixed logit (MMNL) models \citep{McFaddenTrain2000}. The CNL model extends the standard nesting structure by allowing an alternative to belong to multiple nests with fractional allocation parameters, thereby capturing flexible substitution patterns across overlapping groups \citep{VovshaBekhor2002}. The MMNL model, also known as the random parameters logit, introduces random taste heterogeneity by allowing coefficients in the utility function to vary across individuals according to a specified distribution \citep{McFaddenTrain2000,Train2009}. While both models are behaviorally attractive and widely applied, their estimation is significantly more challenging. In the CNL case, the log-likelihood is generally non-concave because the allocation parameters interact nonlinearly with the inclusive values, complicating identification and leading to multiple local optima. Similarly, in the MMNL model,  the likelihood involves high-dimensional integrals over the random coefficients that are approximated through simulation, resulting in a simulated log-likelihood that is also non-concave in the parameters. These non-convexities mean that standard maximum-likelihood solvers may converge slowly or become trapped in local maxima, and that robust and scalable estimation of CNL and MMNL remains substantially more delicate than for MNL or NL.

 The estimation of these discrete choice models has also been implemented in several widely used software tools and libraries. In transportation research, packages such as \texttt{BIOGEME} \citep{Bierlaire2003,Bierlaire2020} and \texttt{Alogit} \citep{Daly1999} are standard, while in economics and marketing, implementations are available in econometric toolboxes such as \texttt{Stata}, \texttt{NLOGIT/LIMDEP} \citep{Greene2016}, and more recently in open-source environments such as \texttt{R} (\texttt{mlogit}, \citealp{Croissant2019}) and \texttt{Python} (\texttt{PyLogit}, \citealp{Brathwaite2018}). Despite differences in interfaces and supported model classes, the underlying estimation routines across these platforms remain largely gradient-based, typically relying on Newton-type or quasi-Newton algorithms to solve the maximum-likelihood problem.

Our work explores a novel estimation approach for the aforementioned discrete choice models by leveraging convex conic optimization.  In this context, it worth mentioning that conic optimization provides a unified framework for solving a broad class of convex programs by expressing them as linear optimization problems over convex cones \citep{BenTalNemirovski2001,BoydVandenberghe2004}. A key advantage of this formulation is that it enables the use of \textit{primal--dual interior-point} methods with \textit{strong polynomial-time complexity guarantees} \citep{NesterovNemirovskii1994}. Within this framework, the exponential cone has emerged as a powerful modeling tool, as it allows nonlinear functions such as the exponential, relative entropy, and the log-sum-exp to be represented exactly in a conic form \citep{Chares2009}. Moreover, casting problems as exponential-cone programs makes them amenable to highly robust and efficient conic solvers such as MOSEK \citep{MOSEKcookbook}, which can handle large-scale instances without the delicate step-size tuning required by first-order methods. Exponential-cone programming underpins applications ranging from geometric programming \citep{BoydGP2007} and logistic regression \citep{McCullaghNelder1989} to entropic optimal transport \citep{Cuturi2013} and distributionally robust optimization \citep{NamkoongDuchi2017}, highlighting its flexibility and wide-ranging impact. While conic optimization, and in particular exponential-cone programming, has been applied in several domains, to the best of our knowledge, this work is the first to employ these techniques for the estimation of discrete choice models.

\section{The MNL Model}\label{sec:MNL}

The multinomial logit (MNL) model is perhaps the most widely used specification in discrete choice analysis \citep{McFadden1974,Train2009}. It assumes that the utility of each alternative consists of a systematic component, linear in observed attributes, plus an unobserved error term following the Type-I extreme value distribution. This assumption yields closed-form choice probabilities, simple behavioral interpretations of parameters, and tractable likelihood functions. Despite its limitations (e.g., the independence of irrelevant alternatives property - IIA), the MNL model remains the workhorse of empirical studies across transportation, marketing, and operations.

\subsection{Maximum Likelihood Estimation}
We first present the standard maximum likelihood estimation of the MNL model. Let $m$ be the number of alternatives, and denote by $[m] = \{1,\ldots,m\}$ the universal set of alternatives. For each individual $i$, the deterministic component of the utility associated with alternative $j \in [m]$ is
$v_{ij} = \bbt^\top \ba_{ij},$
where $\bbt$ is the parameter vector to be estimated, and $\ba_{ij}$ is the vector of observed attributes of alternative $j$ for individual $i$. The full random utility is defined as
$u_{ij} = v_{ij} + \mu \epsilon_{ij},$
where $\epsilon_{ij}$ are i.i.d.\ disturbances following the Type I extreme value (Gumbel) distribution. 
For notational simplicity, we set $\mu = 1$ without loss of generality, since any 
$\mu \neq 1$ can be absorbed into the coefficient vector by defining 
$\tilde{\bbt} = \bbt / \mu$.

Under this distributional assumption, the choice probabilities take the familiar logit form
\[
P_{ij}(S_i) = \frac{\exp(\bbt^\top \ba_{ij})}{\sum_{k \in S_i}\exp(\bbt^\top \ba_{ik})},
\]
where $S_i \subseteq [m]$ denotes the choice set available to individual $i$.

Given $N$ independent observations $(j_n, S_n)$, where $S_n \subseteq [m]$ is the choice set of individual $n$ and $j_n \in S_n$ is the chosen alternative, the maximum likelihood estimator (MLE) solves
\begin{equation}
     \max_{\bbt} \; \left\{\cL^{\MNL}(\bbt) = \sum_{n=1}^N \log \left( \frac{\exp(\bbt^\top \ba_{nj_n})}{\sum_{j\in S_n}\exp(\bbt^\top \ba_{nj})}\right)\right\}.\label{prop:MNL-MLE}\tag{\sf MNL-MLE}
\end{equation}
Equivalently, the log-likelihood can be expressed as
\begin{equation}\label{prob:MNL}
     \cL^{\MNL}(\bbt)  = \sum_{n=1}^N \left[ \bbt^\top \ba_{nj_n} - \log \Big(\sum_{j\in S_n}\exp(\bbt^\top \ba_{nj})\Big)\right],
\end{equation}
which is a concave function problem involving log-sum-exp terms \citep{BoydVandenberghe2004}.

A typical approach to estimation is to compute the gradient of the log-likelihood,
\[
\nabla_{\bbt} \cL^{\MNL}(\bbt)  = \sum_{n=1}^N \left( \ba_{nj_n} - \sum_{j \in S_n} P_{ij}(S_n)\, \ba_{nj} \right),
\]
and then apply a gradient-based routine such as Newton’s method, quasi-Newton methods (e.g., BFGS or L-BFGS), or other iterative schemes \citep{NoceWrig06}. While effective for small- to medium-scale problems, such methods require careful step-size selection and can become computationally demanding when both $N$ and $m$ are large.

\subsection{Exponential Cone Formulation}
We now reformulate the MLE problem under the MNL model as an exponential cone program. 
The key observation is that the log-sum-exp terms in the likelihood can be expressed exactly using the \emph{exponential cone}, a convex set that captures exponential and logarithmic relationships in a conic form.  The exponential cone is typically defined as
\begin{equation}
    \cK_{\exp} \;=\; \left\{ (x_1, x_2, x_3) \;\middle|\; x_2 > 0,\; x_1 \geq x_2 \exp\!\left(\tfrac{x_3}{x_2}\right) \right\}
    \;\cup\; \left\{ (x_1, 0, x_3) \;\middle|\; x_1 \geq 0,\; x_3 \leq 0 \right\}.
\end{equation}
This cone provides a convex representation of nonlinear expressions involving exponentials, logarithms, and relative entropy, and has become a standard tool in modern convex optimization \citep{Chares2009,MOSEKcookbook}. 
\begin{figure}[htb]
    \centering
    \includegraphics[width=0.6\linewidth]{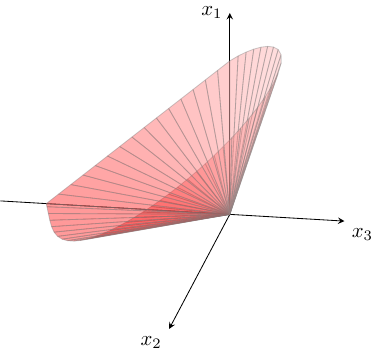}
    \caption{Boundary of the exponential cone $\mathcal{K}_{\exp}$ \citep{MOSEKcookbook}. 
    The surface corresponds to $x_1 = x_2 e^{x_3/x_2}$ for $x_2>0$, while the half-plane at $x_2=0$ captures the limiting case with $x_3 \leq 0$.}
    \label{fig:exp-cone-3d}
\end{figure}

Figure~\ref{fig:exp-cone-3d} illustrates the boundary of the exponential cone. 
The curved surface corresponds to points satisfying $x_1 = x_2 e^{x_3/x_2}$ for $x_2>0$, while the flat region at $x_2=0$ corresponds to the closure of the cone for $x_3 \leq 0$. 
Together, these surfaces delineate the convex region $\cK_{\exp}$ that forms the foundation of our reformulation.

In the context of MLE for the MNL model, the exponential cone provides an exact convex representation for exponential and log-sum-exp functions. 
In particular, problem~\eqref{prob:MNL} can equivalently be written as:
\begin{align}
    \max_{\bbt,\bt}\quad & \sum_{n=1}^N \big(\bbt^\top \ba_{nj_n} - t_n \big) \nonumber\\
    \text{s.t.}\quad 
    & t_n = \log \Big(\sum_{j\in S_n}\exp(\bbt^\top \ba_{nj})\Big), \qquad \forall n \in [N]. \nonumber
\end{align}
Moreover, the equality constraints can be relaxed to the inequality:
\[
t_n \;\geq\; \log \Big(\sum_{j\in S_n}\exp(\bbt^\top \ba_{nj})\Big),
\]
since in the maximization problem the optimizer will always drive $t_n$ to the smallest feasible value, so equality holds at the optimum.  
Rearranging, this inequality is equivalent to
\[
\sum_{j\in S_n} \exp\!\big(\bbt^\top \ba_{nj} - t_n\big) \;\leq\; 1, \qquad \forall n \in [N].
\]
Introducing auxiliary variables $z_{nj} = \exp(\bbt^\top \ba_{nj} - t_n)$, 
the constraints
$\sum_{j\in S_n} \exp\!\big(\bbt^\top \ba_{nj} - t_n\big) \;\leq\; 1 $
can be equivalently written as
\begin{align*}
   & \sum_{j\in S_n} z_{nj} \;\leq\; 1, \qquad \forall n \in [N],\\
   & z_{nj} \;\geq\; \exp\!\big(\bbt^\top \ba_{nj} - t_n\big), 
   \qquad \forall n \in [N], \; j \in S_n.
\end{align*}
The nonlinear inequalities $z_{nj} \geq \exp(\bbt^\top \ba_{nj} - t_n)$ can in turn be expressed 
using exponential cone constraints:
\[
(z_{nj},\, 1,\, \bbt^\top \ba_{nj} - t_n) \;\in\; \cK_{\exp}, 
\qquad \forall n \in [N], \; j \in S_n.
\]
Thus, problem~\eqref{prob:MNL} admits the ECP formulation:
\begin{align}
     \max_{\bbt,\bt,\bz}\quad & \sum_{n=1}^N \big(\bbt^\top \ba_{nj_n} - t_n \big)\label{prop:ECP-MNL}\tag{\sf MNL-ECP} \\
     \text{s.t.}\quad 
     & \sum_{j\in S_n} z_{nj} \leq 1, \qquad \forall n \in [N], \nonumber\\
     & (z_{nj}, 1, \bbt^\top \ba_{nj} - t_n) \in \cK_{\exp}, \qquad \forall n \in [N],\; j \in S_n.\nonumber
\end{align}

\begin{figure}[htb]
\centering
\begin{tikzpicture}[node distance=1cm,>=stealth,thick]
  \node[draw,rounded corners,align=center,
        minimum width=4.5cm,minimum height=1.1cm] (logsum)
    {$t_n \geq \log\!\left(\sum_{j\in S_n} e^{\bbt^\top \ba_{nj}}\right)$};

  \node[draw,rounded corners,align=center,
        minimum width=4.5cm,minimum height=1.1cm,
        right=of logsum] (ineq)
    {$\sum_{j\in S_n} e^{\bbt^\top \ba_{nj} - t_n} \leq 1$};

  \node[draw,rounded corners,align=center,
        minimum width=4.5cm,minimum height=1.1cm,
        right=of ineq] (cone)
    {$(z_{nj}, 1, \bbt^\top \ba_{nj} - t_n) \in \cK_{\exp}$};

  \draw[->] (logsum) -- (ineq);
  \draw[->] (ineq) -- (cone);
\end{tikzpicture}
\caption{\textit{Equivalent reformulations of the log-sum-exp constraint in the MNL likelihood: 
from the nonlinear inequality, to an exponential form, and finally to a conic representation.}}
\label{fig:expcone-MNL}
\end{figure}
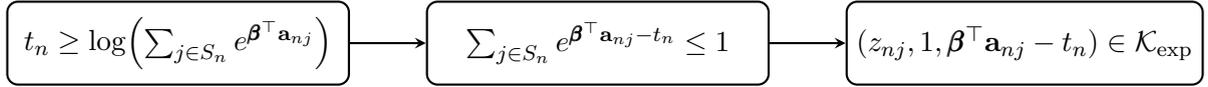
Figure~\ref{fig:expcone-MNL} illustrates the key transformation of the log-sum-exp constraint into its equivalent conic representation.   
The ECP formulation in \eqref{prop:ECP-MNL} is equivalent to the classical MLE but is expressed directly in the language of convex conic optimization. 
As a result, it can be solved efficiently by modern conic solvers (e.g., MOSEK) using interior-point methods, which provide robustness, polynomial-time complexity guarantees \citep{MOSEKcookbook,Nesterov1994interior}.

\subsection{Computational Complexity}
As noted earlier, a key advantage of the ECP reformulation is that it enables the MLE to be solved using \textit{primal--dual interior-point} methods, which enjoy \textit{polynomial-time complexity guarantees} that are difficult to achieve with standard gradient-based approaches. In this section, we analyze the computational complexity of solving the exponential cone program in~\eqref{prop:ECP-MNL}. We begin by examining the size of the program, in terms of its variables and constraints, and then provide a formal statement on the complexity of attaining an $\varepsilon$-optimal solution.

The exponential-cone reformulation in~\eqref{prop:ECP-MNL} introduces a total of $p+N+Z$ decision variables, where $p$ denotes the number of utility coefficients (the dimension of $\bbt$), $N$ is the number of individuals, and $Z=\sum_{n=1}^N |S_n|$ represents the total number of alternative occurrences across all observations.
 These variables correspond respectively to the parameter vector $\bbt$, the auxiliary scalars $t_n$, and the exponential-cone variables $z_{nj}$. The constraints consist of $N$ linear inequalities of the form $\sum_{j\in S_n} z_{nj}\leq 1$ together with $Z$ exponential-cone membership constraints $(z_{nj},1,\bbt^\top \ba_{nj}-t_n)\in\cK_{\exp}$. When expressed in the standard form used by conic solvers, each linear inequality is represented as an equality with an additional nonnegative slack, yielding in total $p+Z+2N$ variables, $N$ linear equalities, and $Z$ exponential-cone blocks of dimension three. In terms of scaling, if $\bar m=\tfrac{1}{N}\sum_n |S_n|$ denotes the average choice-set size, then $Z=N\bar m$, so the problem size grows linearly with both $N$ and $\bar m$, and in the full-availability worst case one has $Z\leq Nm$. 

The following proposition summarizes the overall complexity of solving \eqref{prop:ECP-MNL} via interior-point methods. 

\begin{proposition}[Complexity of solving the MNL-based ECP]
\label{prop:ipm-complexity-mnl-short}
A path-following primal--dual interior-point method applied to the exponential-cone program for \eqref{prop:ECP-MNL} attains an $\varepsilon$-optimal solution in
   $\mathcal{O}\!\Big(\sqrt{Z}\,\log(1/\varepsilon)\;\cdot\;\big(Z\,p^2+(p+N)^3\big)\Big).$
\end{proposition}
The proof (provided in the appendix) follows directly from the standard complexity analysis of interior-point methods 
for self-concordant barriers \citep{NesterovNemirovskii1994,BenTalNemirovski2001}. 
In this framework, the iteration complexity is governed by the barrier parameter $\nu$, which in our case scales as $\nu = \Theta(Z)$ because each exponential-cone block contributes a constant barrier parameter. 
Thus, a path-following method requires $\mathcal{O}(\sqrt{Z}\log(1/\varepsilon))$ Newton steps to obtain an $\varepsilon$-optimal solution. 
The cost per iteration is dominated by assembling and factoring the Schur complement in the global variables $(\bbt,\bt)$, which requires $\mathcal{O}(Zp^2+(p+N)^3)$ arithmetic operations. 
Combining these bounds yields the stated overall complexity. 

The complexity stated in Proposition~\ref{prop:ipm-complexity-mnl-short} implies that the conic reformulation of MNL estimation enjoys the same strong polynomial-time guarantees as other problems in convex optimization. In particular, the number of interior-point iterations grows only with the square root of the total number of exponential-cone blocks $Z$, and only logarithmically with the target accuracy $\varepsilon$. This means that the bulk of the computational burden lies in the per-iteration cost of assembling and solving the Schur complement system, which scales as $\mathcal{O}(Zp^2+(p+N)^3)$. From a practical perspective, this structure is highly favorable: the $\mathcal{O}(Zp^2)$ assembly term is embarrassingly parallel across individuals and alternatives, while the $\mathcal{O}((p+N)^3)$ factorization step depends only on the number of parameters and individuals, not directly on the choice set size. Hence, the conic approach is particularly well suited for large-scale applications where $N$ and the average choice set size are large, and provides robustness and scalability advantages over traditional gradient-based maximum likelihood routines, which lack such complexity guarantees and may suffer from convergence issues in ill-conditioned settings.

\section{The Nested Logit Model}
\label{sec:NL}
The NL model is one of the most widely used extensions of the multinomial logit because of its ability to relax the restrictive IIA property and capture correlations among groups of similar alternatives \citep{Williams1977,DalyZachary1978,Train2009}. Its popularity stems from both its behavioral appeal and its analytical tractability, making it a standard tool in transportation research for mode and route choice analysis, in marketing for product differentiation and market share prediction, and in economics and operations for demand estimation and policy evaluation. Estimation of the NL model typically proceeds by maximum likelihood, and it is well known that when the dissimilarity (scale) parameters are fixed, the log-likelihood is concave in the utility coefficients \citep{DaganzoKusnic1993_convex_nested}. However, joint estimation of the utility and dissimilarity parameters leads to a non-convex problem, raising concerns of local optima and numerical instability \citep{BenAkivaLerman1985,Train2009}. As a result, robust and scalable estimation of the NL model remains more challenging than that of the simpler MNL, despite its broader applicability and flexibility.

\begin{figure}[htb]
\centering
\begin{tikzpicture}[>=stealth,thick]
  \tikzstyle{nest}=[rounded corners,draw,align=center,minimum width=28mm,minimum height=8mm,font=\small]
  \tikzstyle{alt} =[draw,rounded corners=2pt,align=center,minimum width=10mm,minimum height=6mm,font=\small]

  \node[nest] (root) at (0,0) {Root};

  \node[nest] (A) at (-3.6,-1.6) {Nest A\\($\lambda_A$)};
  \node[nest] (B) at ( 3.6,-1.6) {Nest B\\($\lambda_B$)};

  \node[alt] (a1) at (-5.0,-3.2) {$j_1$};
  \node[alt] (a2) at (-2.2,-3.2) {$j_2$};

  \node[alt] (b1) at ( 2.2,-3.2) {$j_3$};
  \node[alt] (b2) at ( 5.0,-3.2) {$j_4$};

  \draw[->] (root.south) -- (A.north);
  \draw[->] (root.south) -- (B.north);

  \draw[->] (A.south) -- (a1.north);
  \draw[->] (A.south) -- (a2.north);

  \draw[->] (B.south) -- (b1.north);
  \draw[->] (B.south) -- (b2.north);

  \node[align=center,font=\footnotesize] at (0,-4.2)
    {For $j \in A$: $P(j)=P(j\mid A)\cdot P(A)$ \quad\quad For $j \in B$: $P(j)=P(j\mid B)\cdot P(B)$};
\end{tikzpicture}
\caption{\textit{Simple nested logit structure with two nests; $\lambda_A,\lambda_B\in(0,1]$ are dissimilarity parameters.}}
\label{fig:nested-structure}
\end{figure}
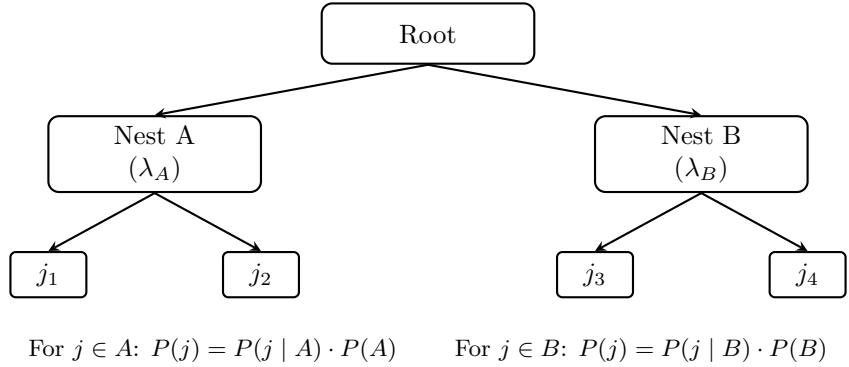

In the NL model, alternatives are partitioned into nests, which form disjoint subsets of the overall choice set. 
The choice probability of each alternative can be decomposed into two components: first, the probability of choosing the nest that contains the alternative, and second, the conditional probability of choosing the alternative within that nest. 
Figure~\ref{fig:nested-structure} illustrates a simple nested structure with two nests, each containing two alternatives. 
For instance, the choice probability of alternative $j_1$ can be expressed as
\[
P(j_1) \;=\; P(\text{Nest A}) \cdot P(j_1 \mid \text{Nest A}),
\]
where $P(\text{Nest A})$ is the probability of selecting Nest~A at the upper level, and $P(j_1 \mid \text{Nest A})$ is the conditional logit probability of choosing $j_1$ among the alternatives within Nest~A. 
Each nest is associated with a dissimilarity (or scale) parameter $\lambda \in (0,1]$, which governs the degree of correlation among the alternatives inside the nest: values of $\lambda$ closer to one indicate weaker correlation (approaching the standard MNL), while smaller values of $\lambda$ allow for stronger correlations within the nest.


\subsection{Maximum Likelihood Estimation}
We first present the MLE of the NL model. 
In the NL model, the full set of alternatives is partitioned into \(L\) disjoint subsets (or nests) \(\cN_1, \ldots, \cN_L \subset [m]\). 
Each nest \(l \in [L]\) is associated with a dissimilarity (scale) parameter \(\lambda_l \in (0,1]\), which measures the degree of independence in unobserved utility among the alternatives within that nest. 

A central feature of the NL model is that the probability of choosing an alternative can be decomposed into two components:  
(i) the probability of selecting the nest that contains the alternative, and  
(ii) the conditional probability of choosing the alternative within that nest. 
Formally, for an observed choice $j_n \in S_n$ made by individual $n$, we can write
\[
P_n(j_n \mid S_n) \;=\; 
P_n(l_n \mid S_n)\,\cdot\,P_n(j_n \mid l_n, S_n),
\]
where $l_n$ denotes the nest containing $j_n$. The probability of choosing nest $l$ is given by
\[
P_n(l \mid S_n) \;=\; 
\frac{W_{nl}^{\lambda_l}}{\sum_{l' \in [L]} W_{nl'}^{\lambda_{l'}}},~~\text{where  }
W_{nl} \;=\; \sum_{j \in \cN_l \cap S_n} 
\exp\!\left(\frac{\bbt^\top \ba_{nj}}{\lambda_l}\right)
\]
is the so-called \emph{inclusive value} (or log-sum term) for nest $l$.  Conditional on nest $l$, the probability of choosing alternative $j \in \cN_l$ is
\[
P_n(j \mid l, S_n) \;=\; 
\frac{\exp\!\left(\bbt^\top \ba_{nj}/\lambda_l\right)}{W_{nl}}.
\]
Combining the two terms, the overall choice probability of $j_n$ can be written as
\[
P_n(j_n \mid S_n) \;=\; 
\frac{W_{n l_n}^{\lambda_{l_n}-1}\,
      \exp\!\left(\bbt^\top \ba_{n j_n}/\lambda_{l_n}\right)}
     {\sum_{l' \in [L]} W_{nl'}^{\lambda_{l'}}}.
\]
The NL model reduces to the standard MNL model when all scale parameters satisfy $\lambda_l = 1$ for every $l \in [L]$, in which case nests collapse to single-level choice sets.

Given the above choice probabilities, the MLE of the NL model can be formulated as  
\begin{align}
    \max_{\bbt,\bld} \; \cL^{\NL}(\bbt,\bld) 
    \;=\; \sum_{n\in [N]} \ln P_n(j_n \mid S_n),
\end{align}
where $P_n(j_n \mid S_n)$ is the nested logit probability of observing choice $j_n$ from the offered set $S_n$ for individual $n$.  Expanding the probability expression, the log-likelihood can be written equivalently as  
\begin{align}
    \max_{\bbt,\bld} \; \cL^{NL}(\bbt,\bld) 
    \;=\; \sum_{n\in [N]} 
    \Bigg[ 
        (\lambda_{l_n}-1)\,\ln W_{n l_n} 
        + \frac{\bbt^\top \ba_{n j_n}}{\lambda_{l_n}}
        - \ln \Bigg(\sum_{l' \in [L]} W_{n l'}^{\lambda_{l'}} \Bigg)\label{prop:NL-MLE} \tag{\sf NL-MLE}
    \Bigg],
\end{align}
The NL log-likelihood has a more complicated structure than that of the MNL model. 
Moreover, it can be shown that if the scale parameters $\bld$ are fixed, then $\cL^{\NL}(\bbt,\bld)$ is concave in $\bbt$ \citep{DaganzoKusnic1993_convex_nested}. 
Standard approaches therefore typically rely on iterative optimization methods that compute the gradient (and sometimes the Hessian) of the log-likelihood with respect to $\bbt$ and update the parameters using Newton or quasi-Newton schemes such as BFGS or L-BFGS. 
When both $\bbt$ and $\bld$ are estimated simultaneously, the problem becomes non-convex, and specialized algorithms or two-step procedures are often employed to improve stability and convergence.

\subsection{ECP Reformulation}
We now present an exponential cone programming (ECP) reformulation of the NL maximum likelihood problem in~\eqref{prop:NL-MLE}. 
The key idea is to replace the nested log-sum-exp and log-sum terms that appear in the likelihood with equivalent exponential-cone constraints, thereby obtaining a tractable conic representation of the estimation problem.
To this end, we introduce auxiliary variables that explicitly represent the inclusive values within each nest and the top-level aggregation across nests. 
Specifically, for individual $n$ and nest $l \in [L]$, define
\begin{align}
    z_{nl}  &= \log\left(\sum_{j\in \cN_l\cap S_n} 
    \exp\!\left(\frac{\bbt^\top \ba_{nj}}{\lambda_l}\right)\right),\\
    y_n  &= \log\left(\sum_{l'\in [L]} W_{nl'}^{\lambda_{l'}}\right),
\end{align}
where $z_{nl}$ represents the log of the inclusive value within nest $l$, and $y_n$ represents the top-level log-sum across all nests.  
With these variables, the NL log-likelihood can be rewritten as the following constrained optimization problem:
\begin{align}
    \max_{\bbt, \{z_{nl}\}, \{y_n\}} \quad & \sum_{n\in [N]} 
    \left[ (\lambda_{l_n}-1) z_{n l_n} + \frac{\bbt^\top \ba_{n j_n}}{\lambda_{l_n}} - y_n \right], \\
    \textbf{s.t. }\quad & z_{nl}  =  \log\!\left(\sum_{j\in \cN_l\cap S_n} 
    \exp\!\left(\frac{\bbt^\top \ba_{nj}}{\lambda_l}\right)\right), \qquad \forall n, l, \label{ctr:NL-1}\\
    & y_n = \log\!\left(\sum_{l'\in [L]}\exp(\lambda_{l'} z_{nl'})\right), \qquad \forall n. \label{ctr:NL-2}
\end{align}
Here we can see that the equality constraints \eqref{ctr:NL-1}--\eqref{ctr:NL-2} can be safely relaxed to inequalities
\begin{align}
    z_{nl}  &\;\geq\;  \log\!\left(\sum_{j\in \cN_l\cap S_n}\nonumber
    \exp\!\left(\frac{\bbt^\top \ba_{nj}}{\lambda_l}\right)\right), ~~\forall n\in [N], l\in [L]\nonumber\\
    y_n &\;\geq\; \log\!\left(\sum_{l'\in [L]} \exp(\lambda_{l'} z_{nl'})\right), ~~\forall n\in [N]\nonumber
\end{align}
without changing the optimal solution. 
This is because $\lambda_l \leq 1$, so in the maximization problem the optimizer will always push $z_{nl}$ and $y_n$ to their smallest feasible values, ensuring the inequalities are tight at optimality.
Next, we rewrite these inequalities in exponential-cone form. Rearranging gives:
\begin{align}
    1 &\;\geq\; \sum_{j\in \cN_l\cap S_n} 
    \exp\!\left(\frac{\bbt^\top \ba_{nj}}{\lambda_l} - z_{nl}\right), ~~\forall n\in [N], l\in [L]\nonumber \\
    1 &\;\geq\; \sum_{l'\in [L]} 
    \exp\!\left(\lambda_{l'} z_{nl'} - y_n\right)  ~~\forall n\in [N].\nonumber
\end{align}
By introducing additional auxiliary variables:
\begin{align}
    k_{njl} &= \exp\!\left(\frac{\bbt^\top \ba_{nj}}{\lambda_l} - z_{nl}\right), \\
    h_{nl'} &= \exp\!\left(\lambda_{l'} z_{nl'} - y_n\right),
\end{align}
the problem can be written as the following exponential cone program:
\begin{align}
   \max_{\bbt, \{z_{nl}\}, \{y_n\}, \{k_{njl}\}, \{h_{nl}\}} \quad 
    & \sum_{n\in [N]} \left[ (\lambda_{l_n}-1) z_{n l_n} + \frac{\bbt^\top \ba_{n j_n}}{\lambda_{l_n}} - y_n \right], \label{prob:NL-ECP}\tag{\sf NL-ECP}\\
    \textbf{s.t. }\quad & \sum_{j\in \cN_l\cap S_n} k_{njl} \;\leq\; 1, 
    \qquad \forall n\in [N], l\in [L], \nonumber\\
    & \sum_{l\in [L]} h_{nl} \;\leq\; 1, 
    \qquad \forall n\in [N],\nonumber \\
    & (k_{njl},\, 1,\, \tfrac{\bbt^\top \ba_{nj}}{\lambda_l} - z_{nl}) \;\in\; \cK_{\exp}, 
    \qquad \forall n\in [N],j\in [m],l\in [L], \\
    & (h_{nl},\, 1,\, \lambda_{l} z_{nl} - y_n) \;\in\; \cK_{\exp}, 
    \qquad \forall n\in [N],l\in [L].
\end{align}

This formulation replaces the  log-sum-exp and log-sum expressions in the NL likelihood with linear and exponential-cone constraints, thereby making the entire estimation problem amenable to modern conic solvers. It preserves the convexity structure when the scale parameters $\{\lambda_l\}$ are fixed and provides a principled way to incorporate NL estimation into the exponential cone programming framework.

\subsection{Computational Complexity}
Similar to the analysis in the previous section, we now discuss the computational complexity of solving \eqref{prob:NL-ECP} using interior-point algorithms. 
For ease of notation, similar to the case of MNL model, let 
\[
Z \;:=\; \sum_{n=1}^N |S_n|,
\qquad
L_n \;:=\; \big|\{\,l\in[L] : \cN_l \cap S_n \neq \emptyset\,\}\big|,
\qquad
\Lambda \;:=\; \sum_{n=1}^N L_n.
\]
Here, $Z$ counts the total number of alternative appearances across all observations, while $\Lambda$ counts the total number of \emph{active} nests across all individuals (i.e., nests that contain at least one available alternative).

In \eqref{prob:NL-ECP}, the main decision variables are as follows:
$\bbt \in \mathbb{R}^p$ ($p$ variables), 
$y_n$ ($N$ variables), 
$z_{nl}$ ($\Lambda$ variables), 
$k_{njl}$ (one for each $(n,j)$ with $j \in \cN_l \cap S_n$, contributing a total of $Z$ variables since nests are disjoint), 
and $h_{nl}$ ($\Lambda$ variables). 
Hence, the model has in total
$p \;+\; N \;+\; Z \;+\; 2\Lambda$
decision variables. The linear constraints consist of:
(i) one inequality $\sum_{j \in \cN_l \cap S_n} k_{njl} \leq 1$ for each active pair $(n,l)$, and 
(ii) one inequality $\sum_{l \in L_n} h_{nl} \leq 1$ for each $n$, 
for a total of $\Lambda + N$ linear inequalities (each of which introduces a nonnegative slack variable in solver-ready form).  

The exponential-cone constraints comprise:  
(i) one block for each $k_{njl}$ ($Z$ blocks), and  
(ii) one block for each $h_{nl}$ ($\Lambda$ blocks),  
so that there are in total
$Z \;+\; \Lambda$
exponential-cone blocks. In the worst case of full availability ($S_n = [m]$ for all $n$) and all nests active for every individual, we have $Z = N m$ and $\Lambda = N L$. 

Given this problem size, the following proposition states the polynomial-time complexity of solving \eqref{prob:NL-ECP} with a path-following interior-point method \citep{Nesterov1994interior}.
\begin{proposition}[Computational  complexity for solving \eqref{prob:NL-ECP}]
\label{prop:ipm-nl-ecp}
A path-following primal--dual interior-point method applied to the NL exponential-cone program \eqref{prob:NL-ECP}
can return  an $\varepsilon$-optimal solution in
$\mathcal{O}\!\Big(\sqrt{Z+\Lambda}\,\log(1/\varepsilon)\ \cdot\ \big(Z\,p^2 + (p+\Lambda+N)^3\big)\Big).$
\end{proposition}
The proof of Proposition \ref{prop:ipm-nl-ecp} can be found in the appendix. The complexity analysis for the NL model highlights both the benefits and the challenges of the exponential cone reformulation. On the one hand, once the scale parameters are fixed, the estimation problem is convex and the inner step can be solved to global optimality with polynomial-time guarantees using interior-point methods. On the other hand, the overall computational cost grows with the total number of observations, the size of the choice sets, and the number of active nests, reflecting the richer structure of the NL compared to the MNL. This scaling underscores why estimation of NL models is substantially more demanding than MNL, and also explains the practical importance of efficient solvers and structural exploitation (e.g., parallel assembly or sparse linear algebra) to handle large-scale applications.


\section{The Tree Nested Logit Model}
\label{sec:TNL}
We discuss the estimation of the TNL model. 
The TNL model extends the standard NL by allowing a hierarchical structure of nests organized as a tree \citep{Train2009, Daly1987}. 
This generalization is particularly useful in applications where alternatives share unobserved components at multiple levels of aggregation, for example in transportation where travelers first choose a travel mode, then a service type within that mode, and finally a specific route, or in marketing where products can be classified by brand, subcategory, and item. 
The tree structure provides greater flexibility than the two-level NL model, as it can capture correlations across alternatives at different depths of the hierarchy. 
However, this added flexibility comes at the cost of increased complexity in estimation. 
In particular, the likelihood involves nested inclusive values at multiple levels of the tree, and while it remains concave in the utility coefficients conditional on fixed scale parameters at each node, the joint estimation of both utilities and scale parameters is non-convex. 
As a result, estimation of TNL models is computationally more demanding and numerically less stable than for standard NL, often requiring multi-stage  procedures to achieve convergence. This complexity has limited the widespread use of TNL relative to simpler logit models, despite its strong behavioral appeal in contexts with naturally hierarchical choice structures. In this section, we show that, similar to the NL case, the hierarchical estimation structure of the TNL model can also be reformulated as an exponential cone program, thereby enabling efficient solution with modern conic optimization methods.

\subsection{Maximum Likelihood Estimation}
In the TNL model, a choice is represented as a path through a tree structure: starting at the root node, the decision-maker selects a sequence of intermediate nodes until reaching a leaf node, which corresponds to the chosen alternative. 
The tree-nested logit thus provides greater flexibility than the standard nested logit, as many sets of alternatives can be naturally organized hierarchically. 
\begin{figure}[htb]
\centering
\begin{tikzpicture}[>=stealth,thick]
  \tikzstyle{nest}=[rounded corners,draw,align=center,minimum width=28mm,minimum height=8mm,font=\small]
  \tikzstyle{alt} =[draw,rounded corners=2pt,align=center,minimum width=10mm,minimum height=6mm,font=\small]

  \node[nest] (root) at (0,0) {Root\\($\lambda_{\text{root}}=1$)};

  \node[nest] (A) at (-4.0,-1.8) {Nest A\\($\lambda_A$)};
  \node[nest] (B) at ( 4.0,-1.8) {Nest B\\($\lambda_B$)};

  \node[nest] (A1) at (-6.0,-3.6) {Nest A1\\($\lambda_{A1}$)};
  \node[nest] (A2) at (-2.0,-3.6) {Nest A2\\($\lambda_{A2}$)};

  \node[nest] (B1) at ( 2.0,-3.6) {Nest B1\\($\lambda_{B1}$)};
  \node[nest] (B2) at ( 6.0,-3.6) {Nest B2\\($\lambda_{B2}$)};

  \node[alt] (a11) at (-6.8,-5.2) {$j_1$};
  \node[alt] (a12) at (-5.2,-5.2) {$j_2$};

  \node[alt] (a21) at (-2.8,-5.2) {$j_3$};
  \node[alt] (a22) at (-1.2,-5.2) {$j_4$};

  \node[alt] (b11) at ( 1.2,-5.2) {$j_5$};
  \node[alt] (b12) at ( 2.8,-5.2) {$j_6$};

  \node[alt] (b21) at ( 5.2,-5.2) {$j_7$};
  \node[alt] (b22) at ( 6.8,-5.2) {$j_8$};

  \draw[->] (root.south) -- (A.north);
  \draw[->] (root.south) -- (B.north);

  \draw[->] (A.south) -- (A1.north);
  \draw[->] (A.south) -- (A2.north);

  \draw[->] (B.south) -- (B1.north);
  \draw[->] (B.south) -- (B2.north);

  \draw[->] (A1.south) -- (a11.north);
  \draw[->] (A1.south) -- (a12.north);

  \draw[->] (A2.south) -- (a21.north);
  \draw[->] (A2.south) -- (a22.north);

  \draw[->] (B1.south) -- (b11.north);
  \draw[->] (B1.south) -- (b12.north);

  \draw[->] (B2.south) -- (b21.north);
  \draw[->] (B2.south) -- (b22.north);

  \node[align=center,font=\footnotesize] at (0,-6.5)
    {$P(j)=\prod_{\text{nodes on path root}\to j} P(\text{child}\mid \text{parent})$;\quad
     internal nodes carry dissimilarity parameters $\lambda\in(0,1]$.};
\end{tikzpicture}
\caption{Three-level tree-nested logit structure with two top-level nests (A and B).}
\label{fig:tree-NL}
\end{figure}
Figure~\ref{fig:tree-NL} illustrates an example of a three-level tree-nested logit structure. 
For example, in the context of transportation mode choice, the hierarchy may take the following form:
\begin{enumerate}
    \item The traveler first decides between private and public transport.
    \item If private transport is chosen, the traveler then selects between a car, motorcycle, or electric vehicle.
    \item If a car is chosen, the traveler finally chooses between a sedan or an SUV.
\end{enumerate}
This hierarchical organization captures correlations at multiple levels of aggregation, allowing for more realistic substitution patterns between alternatives.

We now turn to the mathematical formulation of the TNL model. 
Let the tree have $T$ levels, with the root node at level~1 and internal nodes at levels $1$ through $T-1$ corresponding to nests. 
The leaf nodes at level~$T$ correspond to the set of available alternatives.
Let $\mathcal{N}$ be the set of all nodes and $\mathcal{I}$ be the set of all internal nodes. For any internal node $k \in \mathcal{I}$, let $C(k)$ denote the set of child nodes of $k$. Moreover, let $\mathcal{S}$ be the set of all leaf nodes, which correspond to the alternatives in the choice set. Let $r$ denote the root node of the tree structure.

Each internal node $k \in \mathcal{N}$ in the tree is associated with a scale (or dissimilarity) parameter $\lambda_k > 0$. 
For leaf nodes $k \in \mathcal{S}$, corresponding to the actual alternatives, we normalize $\lambda_k = 1$. 
The choice probability generating function (CPGF) \citep{Fosgerau2013} is then defined recursively as
\begin{equation}
    V_k =  
    \begin{cases}
          \exp(v_k), & \text{if } k \in \mathcal{S}, \\[6pt]
          \displaystyle \sum_{s \in C(k)} V_s^{\lambda_s/\lambda_k}, & \text{if } k \in \mathcal{I},
    \end{cases} 
\end{equation}
where $v_k$ is the deterministic utility of alternative $k$, $\mathcal{I}$ is the set of internal nodes, and $C(k)$ denotes the set of child nodes of $k$. 
To ensure random utility maximization (RUM) consistency, we require $\lambda_s \geq \lambda_k$ for every parent--child pair $(k,s)$ \citep{Train2009}. 
Since $\lambda_s=1$ for all $s \in \mathcal{S}$, this condition implies $\lambda_k \leq 1$ for all $k \in \mathcal{N}$.

The TNL choice probability can be decomposed into conditional probabilities of selecting a child node at each level of the tree. 
Specifically, for any internal node $k \in \mathcal{I}$ and child $s \in C(k)$, the probability of selecting $s$ given $k$ is
\[
P(s \mid k) \;=\; \frac{V_s^{\lambda_s/\lambda_k}}{V_k}.
\]
Thus, the probability of reaching a particular alternative $i \in \mathcal{S}$ can be expressed as the product of conditional probabilities along the unique path from the root $r$ to $i$. 
If this path is denoted $\{k_1 = r, k_2, \ldots, k_T = i\}$, then
\[
P(i \mid \mathcal{S}) \;=\; \prod_{t=1}^{T-1} P(k_{t+1} \mid k_t) 
= \prod_{t=1}^{T-1} \frac{V_{k_{t+1}}^{\lambda_{k_{t+1}}/\lambda_{k_t}}}{V_{k_t}}.
\]
We now describe MLE of the TNL model. 
Suppose we observe $N$ individuals. 
For each individual $n$, the data consists of a pair $(j_n, S_n)$, where $S_n \subseteq [m]$ is the offered choice set and $j_n \in S_n$ is the chosen alternative. 
Let $\{k^n_1 = r, k^n_2, \ldots, k^n_T = j_n\}$ denote the unique path from the root to $j_n$. 
The probability of observing choice $j_n$ is
\begin{align}
    P(j_n \mid S_n) \;=\; \prod_{t=1}^{T-1} \frac{(V^n_{k^n_{t+1}})^{\lambda_{k^n_{t+1}}/\lambda_{k^n_t}}}{V^n_{k^n_t}},
\end{align}
where the values $V^n_k$ are computed recursively as
\[
V^n_k =
\begin{cases}
    \exp(\boldsymbol{\beta}^\top \boldsymbol{a}_{nk}), & \text{if } k \in \mathcal{S}\cap S_n \quad \text{(leaf alternative)}, \\[6pt]
   \displaystyle \sum_{s\in C(k)} (V^n_s)^{\lambda_s/\lambda_k}, & \text{if } k \in \mathcal{I}.
\end{cases}
\]
The log-likelihood for parameters $(\boldsymbol{\beta}, \boldsymbol{\lambda})$ is therefore
\begin{equation}
    \mathcal{L}^{\TNL}(\boldsymbol{\beta}, \boldsymbol{\lambda}) 
    \;=\; \sum_{n=1}^N \ln P(j_n \mid S_n)
    \;=\; \sum_{n=1}^N \sum_{t=1}^{T-1} 
    \left[ \frac{\lambda_{k^n_{t+1}}}{\lambda_{k^n_t}} \ln\!\big(V^n_{k^n_{t+1}}\big) 
    - \ln\!\big(V^n_{k^n_t}\big) \right].
\end{equation}
This formulation highlights how the hierarchical structure of the TNL model decomposes the likelihood into contributions from each level of the decision tree.

\subsection{ECP Reformulation}
We now reformulate the estimation of the TNL model under fixed scale parameters $\{\lambda_k\}_{k\in\cN}$ as an ECP. 
Recall that the choice probability for observation $n$ and chosen alternative $j_n$ can be expressed as
\begin{equation}
   P(j_n \mid S_n) 
   = \prod_{t=1}^{T-1}\frac{(V^n_{k^n_{t+1}})^{\lambda_{k^n_{t+1}}/\lambda_{k^n_t}}}{V^n_{k^n_{t}}} 
   = (V^n_{k^n_1})^{-1}\prod_{t=1}^{T-2} (V^n_{k^n_{t+1}})^{\lambda_{k^n_{t+1}}/\lambda_{k^n_t} - 1}  
     \cdot (V^n_{k^n_{T}})^{\lambda_{k^n_{T}}/\lambda_{k^n_{T-1}}},
\end{equation}
where $\{k^n_1,\ldots,k^n_T=j_n\}$ is the unique path from the root to the chosen leaf node $j_n$. 
Note that $k^n_T$ is a leaf node, hence $\lambda_{k^n_T}=1$ and 
\[
V^n_{k^n_T} = \exp(\bbt^\top \ba_{n j_n}).
\]
The log-likelihood can thus be written as
\begin{align*}
\cL^{\TNL}(\bbt,\bld) 
= \sum_{n\in [N]} \bigg(
    -\ln(V^n_r) 
    + \sum_{t=1}^{T-2}\Big(\tfrac{\lambda_{k^n_{t+1}}}{\lambda_{k^n_t}}-1\Big)\ln V^n_{k^n_{t+1}} 
    + \tfrac{\lambda_{k^n_T}}{\lambda_{k^n_{T-1}}} \, (\bbt^\top \ba_{n j_n})
  \bigg),
\end{align*}
where $V^n_r$ denotes the value function at the root for individual $n$. Recall that 
the value functions $V^n_k$ are defined recursively as
\begin{align}
    V^n_k &= \exp(\bbt^\top \ba_{nk}), 
    && k \in \cS \cap S_n, \quad \text{(leaf alternative)} \label{eq:TNL-leaf}\\
    V^n_k &= \sum_{s \in C(k)} (V^n_s)^{\lambda_s/\lambda_k}, 
    && k \in \cI, \quad \text{(internal node)}. \label{eq:TNL-internal}
\end{align}
Because $\lambda_s \geq \lambda_k$ for all parent--child pairs $(k,s)$, the coefficients of $\ln V^n_k$ in the log-likelihood are non-positive.  Thus, in maximizing the log-likelihood, the optimizer will push $V^n_k$ to be as small as possible, forcing tightness at the optimum. Therefore, the equalities in \eqref{eq:TNL-leaf}--\eqref{eq:TNL-internal} can be relaxed to inequalities:
\begin{align*}
  V^n_k &\;\geq\; \exp(\bbt^\top \ba_{nk}), && k \in \cS\cap S_n, \;\; n \in [N],\\
  V^n_k &\;\geq\; \sum_{s\in C(k)} (V^n_s)^{\lambda_s/\lambda_k}, && k \in \cI, \;\; n \in [N].
\end{align*}
We now let $z^n_k = \ln V^n_k$ for all $k \in \cN$.  The problem can then be reformulated as
\begin{align}
    \max_{\bbt, \bz} \quad & \sum_{n\in [N]} \left(
        -z^n_r 
        + \sum_{t=1}^{T-2}\Big(\tfrac{\lambda_{k^n_{t+1}}}{\lambda_{k^n_t}}-1\Big) z^n_{k^n_{t+1}} 
        + \tfrac{\lambda_{k^n_T}}{\lambda_{k^n_{T-1}}} (\bbt^\top \ba_{n j_n})
    \right)\label{prob:TNL-log}\\
    \text{s.t.} \quad & z^n_k \;\geq\; \bbt^\top \ba_{nk}, && k \in \cS\cap S_n, \; n\in [N],\\
    & \exp(z^n_k) \;\geq\; \sum_{s\in C(k)} \exp\!\Big(\tfrac{\lambda_s}{\lambda_k} z^n_s\Big),
    && k \in \cI, \; n \in [N]. \label{ctr:TNL-exp}
\end{align}
Constraint \eqref{ctr:TNL-exp} can be rewritten as
\[
1 \;\geq\; \sum_{s\in C(k)} \exp\!\left(\tfrac{\lambda_s}{\lambda_k} z^n_s - z^n_k\right), 
\qquad k \in \cI,\; n\in[N].
\]
We introduce auxiliary variables
\[
y^n_{ks} = \exp\!\left(\tfrac{\lambda_s}{\lambda_k} z^n_s - z^n_k\right), 
\qquad \forall k\in \cI, \; s \in C(k), \; n\in [N],
\]
so that the problem becomes the following exponential cone program:
\begin{align}
    \max_{\bbt, \bz, \by} \quad & \sum_{n\in [N]} \left(
        -z^n_r 
        + \sum_{t=1}^{T-2}\Big(\tfrac{\lambda_{k^n_{t+1}}}{\lambda_{k^n_t}}-1\Big) z^n_{k^n_{t+1}} 
        + \tfrac{\lambda_{k^n_T}}{\lambda_{k^n_{T-1}}} (\bbt^\top \ba_{n j_n})
    \right)\tag{\sf TNL-ECP}\label{prob:TNL-ECP}\\
    \text{s.t.}\quad & z^n_k \;\geq\; \bbt^\top \ba_{nk}, && k \in \cS\cap S_n, \; n\in[N],\\
    & 1 \;\geq\; \sum_{s\in C(k)} y^n_{ks}, && k \in \cI, \; n\in[N],\\
    & (y^n_{ks},\,1,\, \tfrac{\lambda_s}{\lambda_k} z^n_s - z^n_k) \;\in\; \cK_{\exp}, 
    && k \in \cI, \; s \in C(k), \; n \in [N].
\end{align}

This ECP formulation explicitly replaces the nested log-sum-exp constraints of the TNL likelihood with linear inequalities and exponential cone constraints, making the problem amenable to modern interior-point conic solvers (when all the scale parameters $\lambda_k$ are fixed).

\subsection{Computational Complexity}
We now discuss the computational complexity of solving \eqref{prob:TNL-ECP} (with fixed scale parameters) using interior-point methods. 
Analogous to the analysis for the MNL and NL models, we begin by defining the aggregate counts:
\[
Z \;:=\; \sum_{n=1}^N |S_n|, 
\qquad 
A_n \;:=\; \{\,k\in\cI : \text{the subtree of }k\text{ contains some } s\in S_n\,\},
\qquad 
\Gamma \;:=\; \sum_{n=1}^N |A_n|.
\]
Thus, $Z$ is the total number of leaf appearances across all observations and $\Gamma$ is the total number of \emph{active} internal nodes across individuals (i.e., internal nodes whose subtrees intersect the offered set).
Let $E_n$ be the set of active parent--child edges in the minimal subtree that spans $S_n$ and the root; then $|E_n| = |A_n|+|S_n|-1$, and the total number of active edges across observations is
\[
E \;:=\; \sum_{n=1}^N |E_n| \;=\; \Gamma + Z - N .
\]
In the TNL-ECP \eqref{prob:TNL-ECP}, the decision variables  are:
$\bbt\in\bbR^p$ ($p$ vars), node logs $\{z_k^n\}$ for all active nodes ($\Gamma+Z$ vars), and edge auxiliaries $\{y^n_{ks}\}$ for all active edges ($E$ vars).
Hence the model has
\[
 p \;+\; (\Gamma+Z) \;+\; E .
\]
variables. The \emph{linear} inequalities are one constraint $1\ge \sum_{s\in C(k)} y^n_{ks}$ for each active internal node $k\in A_n$ (total $\Gamma$ constraints, adding $\Gamma$ nonnegative slacks in solver form) and the affine leaf bounds $z_k^n\ge \bbt^\top \ba_{nk}$ for each $(n,k)$ with $k\in S_n$ (total $Z$ affine rows).
The \emph{exponential-cone} blocks consist of one block per active edge $(k,s)\in E_n$, so there are $E$ three-dimensional $\cK_{\exp}$ blocks in total.
In the worst case of full availability ($S_n=\cS$ for all $n$) on a tree with $m:=|\cS|$ leaves and $q:=|\cI|$ internal nodes, we have $\Gamma=Nq$ and $E=N(q+m-1)=\Theta(Nm)$.

The following proposition states the computational complexity of solving \eqref{prob:TNL-ECP} using a path-following interior-point algorithm \citep{NesterovNemirovskii1994}.

\begin{proposition}[Complexity of solving the TNL-based ECP]
\label{prop:ipm-tnl-ecp}
Consider the TNL exponential-cone program \eqref{prob:TNL-ECP} with fixed scale parameters $\{\lambda_k\}$. The path-following primal--dual interior-point method computes an $\varepsilon$-optimal solution in
${\;
\mathcal{O}\!\Big(\sqrt{E}\,\log(1/\varepsilon)\ \cdot\ \big(Z\,p^2 + (p+\Gamma+Z)^3\big)\Big)
\;}$
\end{proposition}
The proof can be found in the appendix. The complexity formulation for the TNL model underscores the additional computational burden introduced by the hierarchical tree structure. Compared to the NL model, the per-iteration cost of the interior-point method now depends not only on the number of observations and choice set sizes, but also on the number of active internal nodes and edges in the decision tree. This reflects the richer substitution patterns captured by TNL, but also implies that estimation can become considerably more expensive for deep or wide trees. At the same time, the analysis shows that the problem remains polynomial-time solvable when scale parameters are fixed.

\section{Estimation Methods}
\label{sec:method}
In this section, we present a general framework for estimating the MNL, NL, and TNL models, building on their ECP reformulations. 
When the scale parameters are fixed, as in the MNL model and in the NL and TNL models under fixed $\blambda$, the estimation problem is convex and can be solved efficiently to global optimality using off-the-shelf conic solvers such as MOSEK \citep{MOSEKcookbook}.

In the more general case where the scale parameters $\{\blambda\}$ must be estimated jointly with the utility coefficients $\bbt$, the problem becomes non-convex due to the nonlinear dependence of the likelihood on $\blambda$. 
To address this, we adopt a two-stepstep  procedure that alternates between updating the scale parameters and optimizing the utility parameters:
\begin{itemize}
    \item \textbf{Outer step (updating $\bld$):} Given a current estimate of the utility coefficients $\bbt$, update the scale parameters $\{\blambda\}$ by solving the restricted likelihood problem with $\bbt$ fixed at optimum. This step may be performed using constrained nonlinear optimization methods (e.g., projected gradient or trust-region methods) under the admissibility constraints $0<\lambda_k \leq 1$ and $\lambda_s \geq \lambda_k$ for each parent--child pair $(k,s)$. 
    
    \item \textbf{Inner step (optimizing $\bbt$):} Given current values of the scale parameters $\{\blambda\}$, optimize the utility coefficients $\bbt$ by solving the corresponding ECP using a conic solver. Since this subproblem is convex, the inner step yields a globally optimal update for $\bbt$.
\end{itemize}
For the outer update step, the gradient with respect to the parameter $\blambda$ can be obtained by differentiating the log-likelihood function while holding $\bbt$ fixed at its optimal value. Specifically, let  
   $ \bbt^*(\blambda) = \argmax_{\bbt} \; \cL^{\NL}(\bbt, \blambda),$
and define  
  $  \cL^{\NL*}(\blambda) = \cL^{\NL}(\bbt^*(\blambda), \blambda).$
According to the envelope theorem~\citep{milgrom2002envelope}, the gradient of $\cL^{\NL*}(\blambda)$ with respect to $\blambda$ can then be expressed as  
\begin{equation}\label{}
    \nabla_{\blambda} \cL^{\NL*}(\blambda) 
    = \nabla_{\blambda} \cL^{\NL}(\bbt^*(\blambda), \blambda).
\end{equation}
A similar gradient formulation can be derived for the TNL model. These gradients admit a closed-form expression and can be supplied to a nonlinear constrained optimization solver to efficiently solve the outer problem.

\begin{figure}[htb]
\centering
\begin{tikzpicture}[node distance=1cm,>=stealth,thick]

  \tikzstyle{block} = [draw, rounded corners, align=center,
        minimum width=3cm, minimum height=1.0cm]
  \tikzstyle{decision} = [diamond, aspect=2, draw, align=center,
        inner sep=1pt, minimum width=4.5cm, minimum height=1.3cm]

  \node[block] (initAll) {Set $\epsilon>0$ as optimal gap \\
     Set $\mathcal{M}_E$ as ECP model (with fixed $\bld$)\\
     Set $\mathcal{M}_S$ as the MLE objective function with fixed $\bbt$\\
     Initialize $\{\lambda_{k}\}$ based on their bounds};

  \node[decision, below=of initAll] (cond)
    {MLE objective increment $< \epsilon$ \\ or runtime limit reached?};

  \node[block, below=1cm of cond] (step1)
    {Build $\mathcal{M}_E$ with given $\lambda$ and solve \\
    by an ECP solver (e.g. MOSEK) $\;\to\;$ get new $\bbt$};

  \node[block, below=of step1] (step2)
    {Build $\mathcal{M}_S$ with given $\beta$ and solve \\ by a constrained optimization solver (e.g., Scipy) $\;\to\;$ get new  $\bld$};

  \node[block, right=1cm of cond] (ret)
    {Return $(\bbt, \bld)$};

  \draw[->] (initAll) -- (cond);
  \draw[->] (cond.south) -- ++(0,-0.7) -- (step1.north) node[midway,right,yshift=0.5cm] {No};
  \draw[->] (step1) -- (step2);


 \draw[->, rounded corners] 
    (step2.south) -- ++(0,-1) -- ++(-6,0) -- ++(0,6.95) -- ++(2,0) -- (cond.west);

  \draw[->] (cond.east) -- (ret.west) node[midway,above,xshift=-0.1cm] {Yes};

\end{tikzpicture}
\caption{Iterative estimation procedure for estimating NL and  TNL models.}
\label{fig:IOP}
\end{figure}
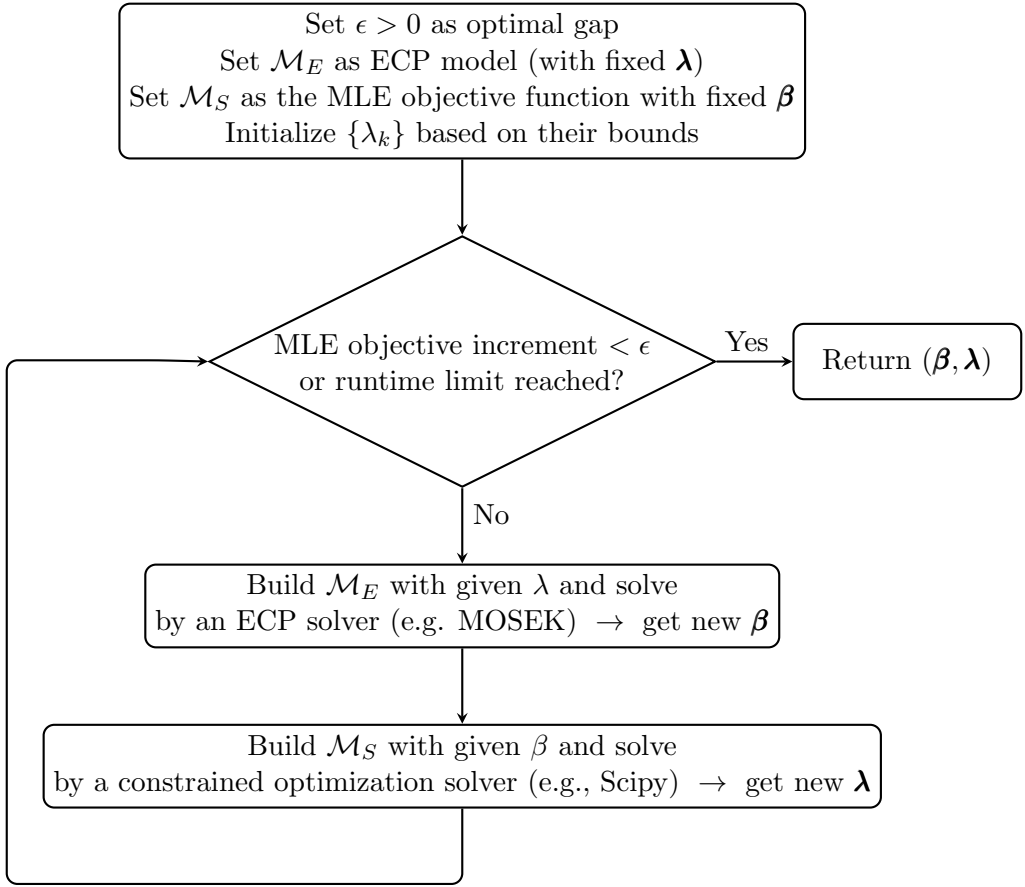

Figure~\ref{fig:IOP} illustrates the pseudo-code of our two-step procedure.
This alternating two-step procedure decouples the non-convex estimation problem into a sequence of convex inner problems and constrained nonlinear outer updates. 
The inner conic programs leverage the computational advantages of exponential cone reformulations, while the outer updates adjust the scale parameters under their structural constraints. 
In practice, we iterate between the outer and inner steps until convergence of the log-likelihood or until successive updates fall below a prescribed tolerance. 
This framework provides a flexible and scalable estimation method applicable to the MNL, NL, and TNL models, encompassing both fixed and estimated scale parameter settings.



\section{Numerical Experiments}
\label{sec:results}
We conduct experiments to evaluate the performance of our conic optimization approach in comparison with standard gradient-based methods. The experiments are based on generated model structures and datasets of varying sizes, with the objective of assessing how the ECP reformulation performs, particularly in large-scale instances.

\subsection{Data Generation \& Experimental Setting}
To examine the performance of the ECP reformulations, we randomly generate three sets of instances corresponding to the MNL, NL, and TNL models.

\paragraph{MNL instances.} We generate $36$ sets of MNL instances with parameter configurations 
$\text{dim}(\bbt) \in \{5,10,20,50\}$, 
$(N,m) \in \{(500,50), (1000,100), (2000,200)\}$, 
and $|S_n| \in \{0.2m, 0.5m, 0.8m\}$. 
Each set consists of $5$ instances. For each instance, every element of the vector 
$\mathbf{a}_{ij}$ ($i \in N, j \in [m]$) is drawn independently from $U[0,3]$. 
The choice set $S_n$ and the chosen alternative $j_n$ are randomly selected from $[m]$, 
subject to the condition that $j_n \in S_n$. 

\paragraph{NL instances.} For the NL formulation, we generate $72$ sets of instances using the same specifications 
for $\dim(\bbt)$, $(N,m)$, $|S_n|$, and $\mathbf{a}_{ij}$ as in the MNL case. 
In addition, the number of nests is set to $L \in \{2,5\}$. 

\paragraph{TNL instances.} 
For the TNL model, we also generate $72$ sets of instances. 
Here we consider a tree structure with depth $T=4$, 
where the number of child nodes at levels $1$ and $2$ is chosen from $\{2,3\}$. 

For cases where $\blambda$ is fixed, we draw the upper bound $u_{\lambda}$ from $U[0.8,0.9]$, 
the lower bound $l_{\lambda}$ from $U[0.1,0.2]$, 
and generated each $\lambda$ uniformly from the interval $[l_{\lambda}, u_{\lambda}]$.

We compare our approach with the nonlinear optimization solvers implemented in 
\texttt{SciPy.optimize}---a state-of-the-art library that provides gradient-based methods 
for minimizing (or maximizing) continuous nonlinear objective functions, possibly subject 
to constraints \citep{2020SciPy-NMeth}. The package includes solvers for general nonlinear problems and supports 
both local and global optimization algorithms.

To select the most suitable solver from the \texttt{optimize} module, we conducted 
a preliminary experiment on several instances of the MNL, NL, and TNL datasets with 
five attributes. The results (reported in the Appendix) indicate that \texttt{L-BFGS-B} 
and \texttt{SLSQP} achieve the lowest runtime, number of iterations (\texttt{nit}), 
and number of function evaluations (\texttt{nfev}) among the solvers available in 
\texttt{SciPy.optimize}. Consequently, \texttt{L-BFGS-B} is selected as the baseline 
solver for estimating the MNL, NL, and TNL models with fixed $\bld$, while \texttt{SLSQP} is employed 
for estimating the NL and TNL models (jointly estimating $\bbt$ and $\bld$) due to its ability to handle constraints on $\blambda$. 

All methods in our experiments are implemented in Python, using 
\texttt{MOSEK 11.0} and \texttt{SciPy 1.15.2}, with a solving time limit of one hour per run. 
The experiments are conducted on a PC equipped with an Intel(R) Core(TM) i7-9700 CPU 
@ 3.00GHz, 16 GB of RAM, and the Windows~11 operating system.

\subsection{Comparison Results}
We present comparative results between our ECP-based method and gradient-based 
baselines across various estimation instances of the MNL, NL, and TNL models, 
considering both the fixed $\blambda$ case and the joint estimation of 
$(\bbt, \blambda)$.

\subsubsection{Estimation of MNL, NL and TNL Models with Fixed $\bld$}
\begin{table}[htb]\footnotesize
\centering
\resizebox{\textwidth}{!}{%
\begin{tabular}{cclrrlrrlrrlrrlrrlrr}
\toprule
 &
   &
   &
  \multicolumn{5}{c}{MNL} &
   &
  \multicolumn{5}{c}{NL} &
   &
  \multicolumn{5}{c}{TNL} \\ \cmidrule{4-8} \cmidrule{10-14} \cmidrule{16-20} 
 &
   &
   &
  \multicolumn{2}{c}{L-BFGS-B} &
   &
  \multicolumn{2}{c}{ECP} &
   &
  \multicolumn{2}{c}{L-BFGS-B} &
   &
  \multicolumn{2}{c}{ECP} &
   &
  \multicolumn{2}{c}{L-BFGS-B} &
   &
  \multicolumn{2}{c}{ECP} \\ \cmidrule{4-5} \cmidrule{7-8} \cmidrule{10-11} \cmidrule{13-14} \cmidrule{16-17} \cmidrule{19-20} 
\#Att &
  Size &
   &
  \#Opt &
  AveTime(s) &
   &
  \#Opt &
  AveTime(s) &
   &
  \#Opt &
  AveTime(s) &
   &
  \#Opt &
  AveTime(s) &
   &
  \#Opt &
  AveTime(s) &
   &
  \#Opt &
  AveTime(s) \\ \midrule
\multirow{3}{*}{5} &
  S &
   &
  \textbf{15} &
  1.48 &
   &
  \textbf{15} &
  \textbf{0.47} &
   &
  26 &
  7.14 &
   &
  \textbf{30} &
  \textbf{0.94} &
   &
  27 &
  5.79 &
   &
  \textbf{30} &
  \textbf{0.17} \\
 &
  M &
   &
  \textbf{15} &
  5.79 &
   &
  \textbf{15} &
  \textbf{1.60} &
   &
  26 &
  33.79 &
   &
  \textbf{30} &
  \textbf{3.66} &
   &
  27 &
  19.5 &
   &
  \textbf{30} &
  \textbf{0.62} \\
 &
  L &
   &
  \textbf{15} &
  23.61 &
   &
  \textbf{15} &
  \textbf{11.65} &
   &
  24 &
  144.24 &
   &
  \textbf{30} &
  \textbf{13.22} &
   &
  26 &
  103.64 &
   &
  \textbf{30} &
  \textbf{2.71} \\ \midrule
\multirow{3}{*}{10} &
  S &
   &
  \textbf{15} &
  2.14 &
   &
  \textbf{15} &
  \textbf{0.58} &
   &
  24 &
  21.17 &
   &
  \textbf{30} &
  \textbf{0.85} &
   &
  29 &
  14.9 &
   &
  \textbf{30} &
  \textbf{0.20} \\
 &
  M &
   &
  \textbf{15} &
  8.41 &
   &
  \textbf{15} &
  \textbf{2.15} &
   &
  20 &
  85.91 &
   &
  \textbf{30} &
  \textbf{3.03} &
   &
  26 &
  60.2 &
   &
  \textbf{30} &
  \textbf{0.70} \\
 &
  L &
   &
  \textbf{15} &
  32.63 &
   &
  \textbf{15} &
  \textbf{11.25} &
   &
  21 &
  412.45 &
   &
  \textbf{30} &
  \textbf{16.45} &
   &
  23 &
  261.33 &
   &
  \textbf{30} &
  \textbf{2.78} \\ \midrule
\multirow{3}{*}{20} &
  S &
   &
  12 &
  3.81 &
   &
  \textbf{15} &
  \textbf{0.59} &
   &
  7 &
  51.26 &
   &
  \textbf{30} &
  \textbf{1.20} &
   &
  22 &
  62.86 &
   &
  \textbf{30} &
  \textbf{0.25} \\
 &
  M &
   &
  10 &
  12.64 &
   &
  \textbf{15} &
  \textbf{2.50} &
   &
  8 &
  202.43 &
   &
  \textbf{30} &
  \textbf{4.33} &
   &
  11 &
  294.12 &
   &
  \textbf{30} &
  \textbf{0.83} \\
 &
  L &
   &
  13 &
  56.74 &
   &
  \textbf{15} &
  \textbf{12.44} &
   &
  7 &
  1079.89 &
   &
  \textbf{30} &
  \textbf{20.31} &
   &
  22 &
  781.98 &
   &
  \textbf{30} &
  \textbf{3.21} \\ \midrule
\multirow{3}{*}{50} &
  S &
   &
  13 &
  11.26 &
   &
  \textbf{15} &
  \textbf{1.24} &
   &
  4 &
  332.15 &
   &
  \textbf{30} &
  \textbf{1.69} &
   &
  8 &
  449.16 &
   &
  \textbf{30} &
  \textbf{0.41} \\
 &
  M &
   &
  \textbf{15} &
  44.56 &
   &
  \textbf{15} &
  \textbf{3.54} &
   &
  9 &
  1686.23 &
   &
  \textbf{30} &
  \textbf{11.14} &
   &
  20 &
  1100.22 &
   &
  \textbf{30} &
  \textbf{1.33} \\
 &
  L &
   &
  \textbf{15} &
  165.42 &
   &
  \textbf{15} &
  \textbf{16.71} &
   &
  1 &
  3372.16 &
   &
  \textbf{30} &
  \textbf{26.95} &
   &
  15 &
  2820.66 &
   &
  \textbf{30} &
  \textbf{6.39} \\ \midrule
\multicolumn{2}{r}{Summary:} &
   &
  168 &
   &
   &
  \textbf{180} &
   &
   &
  177 &
   &
   &
  \textbf{360} &
   &
   &
  256 &
   &
   &
  \textbf{360} &
   \\ \bottomrule
\end{tabular}%
}
\caption{Comparison results for the MNL, NL and TNL instances (with fixed $\bld$).}
\label{tab:MNL_NL_TreeNL}
\end{table}

Table~\ref{tab:MNL_NL_TreeNL} compares the performance of \texttt{L-BFGS-B} and 
the proposed ECP solver across the MNL, NL, and TNL models (with fixed $\bld$). The instances are grouped by numbers of attributes, $\{5,10,20,50\}$ (denoted by ``\#Att''), and dataset sizes 
$(N,m) \in \{(500,50), (1000,100), (2000,200)\}$ (denoted by $\{S, M, L\}$). The detailed results of each set in 36 sets are visualized in the Appendix \ref{append:choice_set}. 
In the table, the columns ``\#Opt'' report the number of optimal solutions found by each 
method for a given set of instances, while the columns ``AveTime(s)'' present the average 
solving time in seconds. The best results in each row are highlighted in bold. 

The results demonstrate that ECP consistently outperforms \texttt{L-BFGS-B} in both 
solution reliability and computational efficiency. Specifically,
for the {MNL instances,} across all $180$ instances, ECP achieves a perfect optimization 
    rate of $180/180$, compared to $168/180$ for \texttt{L-BFGS-B}.  
For the {NL instances,} ECP solves all $360$ instances successfully, whereas 
    \texttt{L-BFGS-B} solves only $177$.  
For the  {TNL instance,} ECP again achieves a perfect success rate of $360/360$, while 
    \texttt{L-BFGS-B} solves $256$ instances.  In terms of runtime, ECP is significantly faster. In the NL model with $50$ attributes and large dataset size ($L$), 
    ECP requires on average only $26.95$ seconds, compared to $3372.16$ seconds for 
    \texttt{L-BFGS-B}.  
 In the TNL model under the same configuration, ECP completes in $6.39$ seconds, 
    while \texttt{L-BFGS-B} requires $2820.66$ seconds.  

Overall, these results highlight the superior scalability and robustness of ECP, 
establishing it as the preferred method for efficiently solving large-scale 
and nested logit models.

\begin{figure}[htb]
    \centering
    \includegraphics[width=0.45\linewidth]{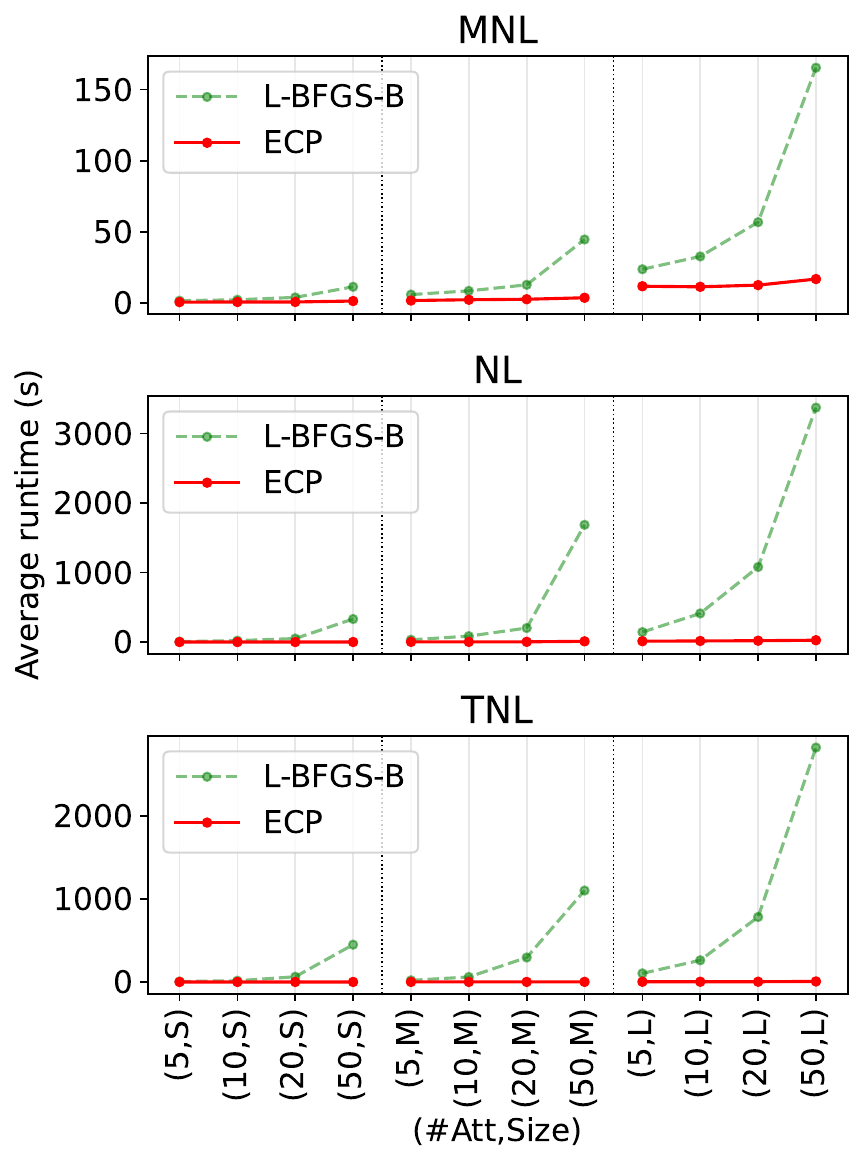}
    \caption{Comparison of average solving time across different methods on the MNL, NL and TNL datasets.}
    \label{fig:avetime_size}
\end{figure}

Figures~\ref{fig:avetime_size} illustrate the average solution times of the 
optimization methods for estimating MNL, NL, and TNL instances (with fixed $\bbt$). The results show 
that \texttt{ECP} consistently outperforms \texttt{L-BFGS-B}, particularly on 
medium and large instances, where the runtime of \texttt{L-BFGS-B} increases sharply.

\subsubsection{Joint Estimating $(\bbt,\bld)$ for the  NL Model}
Table~\ref{tab:GeneralNL} presents the comparative performance of three estimation 
approaches---\texttt{L-BFGS-B}, \texttt{Mixed L-BFGS-B}, and \texttt{ECP+L-BFGS-B}---for 
estimating the NL model across varying sizes of the parameter vector $\bbt$ 
(column ``\#Att''), the number of nests  (column ``\#Nest''), 
and dataset sizes (see Appendix \ref{append:choice_set} for more details of each set in 72 sets). For the \texttt{L-BFGS-B} baseline, we directly apply the solver to estimate the 
MLE. 
The \texttt{Mixed L-BFGS-B} baseline adapts our two-stage procedure 
(Section~\ref{sec:method}), where both stages are solved using \texttt{L-BFGS-B}. 
Finally, our proposed method, \texttt{ECP+L-BFGS-B}, employs \texttt{L-BFGS-B} 
in the outer step and the ECP solver in the inner step to solve the ECP 
reformulations.
Because global optimality cannot be guaranteed for the joint estimation problems, 
the columns ``\#Opt'' are replaced by ``\#Best,'' which report the number of best 
solutions found. In addition, the columns ``AveGap(\%)'' present the average 
objective gaps of \texttt{Mixed L-BFGS-B} and \texttt{ECP+L-BFGS-B} relative 
to \texttt{L-BFGS-B}. 

The results show that our approach \texttt{ECP+L-BFGS-B} approach consistently 
achieves the highest number of best solutions ($342/360$), 
far surpassing \texttt{Mixed L-BFGS-B} ($262/360$) and \texttt{L-BFGS-B} alone ($136/360$). 
Moreover, \texttt{ECP+L-BFGS-B} not only attains superior solution quality but also 
offers faster computation times in nearly all settings. For example, 
with $50$ attributes, $5$ nests, and a large dataset, \texttt{ECP+L-BFGS-B} finds 
all $15$ best solutions in an average of $1205.79$ seconds, compared to 
$3600.00$ seconds for \texttt{Mixed L-BFGS-B} (which finds only one best solution) 
and $3254.23$ seconds for \texttt{L-BFGS-B} (which finds only two). 

In terms of average gaps, \texttt{ECP+L-BFGS-B} again demonstrates a clear advantage, 
particularly in high-dimensional settings. For instance, with $50$ attributes and 
large datasets, it achieves average gaps of $70.61\%$ and $56.00\%$, 
significantly outperforming \texttt{Mixed L-BFGS-B}, which fails to yield meaningful 
gaps (reported as $-\infty$). 

Overall, these results demonstrate that our approach provides 
a powerful and scalable solution for the NL estimation problem, excelling in both 
efficiency and reliability.

\begin{table}[!h]\footnotesize
\centering
\resizebox{\textwidth}{!}{%
\begin{tabular}{ccclrrlrrrlrrr}
\toprule
 &
   &
   &
   &
  \multicolumn{2}{c}{L-BFGS-B} &
   &
  \multicolumn{3}{c}{Mixed L-BFGS-B} &
   &
  \multicolumn{3}{c}{ECP+L-BFGS-B} \\ \cmidrule{5-6} \cmidrule{8-10} \cmidrule{12-14} 
\#Att &
  \#Nest &
  Size &
   &
  \#Best &
  AveTime(s) &
   &
  \#Best &
  AveTime(s) &
  AveGap(\%) &
   &
  \#Best &
  AveTime(s) &
  AveGap(\%) \\ \midrule
\multirow{6}{*}{5} &
  \multirow{3}{*}{2} &
  S &
   &
  5 &
  10.24 &
   &
  \textbf{14} &
  24.92 &
  17.54 &
   &
  \textbf{14} &
  \textbf{8.90} &
  17.54 \\
        &         & M        &  & 6   & 44.32          &  & 13          & 92.11   & 3.04   &  & \textbf{14}  & \textbf{36.58}   & 3.04  \\
        &         & L        &  & 3   & 181.31         &  & 9           & 543.87  & 21.37  &  & \textbf{14}  & \textbf{143.29}  & 21.37 \\ \cmidrule{2-14}
 &
  \multirow{3}{*}{5} &
  S &
   &
  10 &
  14.19 &
   &
  \textbf{15} &
  19.75 &
  7.93 &
   &
  \textbf{15} &
  \textbf{12.45} &
  7.93 \\
        &         & M        &  & 8   & \textbf{47.97} &  & \textbf{15} & 91.69   & 7.88   &  & 14           & 59.03            & 7.88  \\
        &         & L        &  & 7   & 229.20         &  & 13          & 449.13  & 11.34  &  & \textbf{15}  & \textbf{206.11}  & 11.34 \\ \midrule
\multirow{6}{*}{10} &
  \multirow{3}{*}{2} &
  S &
   &
  8 &
  23.85 &
   &
  12 &
  61.79 &
  9.69 &
   &
  \textbf{14} &
  \textbf{11.30} &
  9.69 \\
        &         & M        &  & 6   & 99.23          &  & 9           & 327.87  & 16.91  &  & \textbf{15}  & \textbf{52.11}   & 16.91 \\
        &         & L        &  & 3   & 496.86         &  & 10          & 1432.69 & 20.87  &  & \textbf{13}  & \textbf{207.52}  & 20.87 \\ \cmidrule{2-14}
 &
  \multirow{3}{*}{5} &
  S &
   &
  10 &
  \textbf{26.95} &
   &
  12 &
  58.61 &
  9.06 &
   &
  \textbf{14} &
  28.69 &
  9.06 \\
        &         & M        &  & 9   & 94.60          &  & \textbf{15} & 214.97  & 12.49  &  & \textbf{15}  & \textbf{76.18}   & 12.49 \\
        &         & L        &  & 5   & 373.83         &  & \textbf{14} & 884.80  & 12.99  &  & \textbf{14}  & \textbf{351.61}  & 12.99 \\ \midrule
\multirow{6}{*}{20} &
  \multirow{3}{*}{2} &
  S &
   &
  3 &
  58.55 &
   &
  10 &
  247.40 &
  49.19 &
   &
  \textbf{14} &
  \textbf{35.40} &
  49.19 \\
        &         & M        &  & 5   & 296.79         &  & 12          & 1124.31 & 30.58  &  & \textbf{13}  & \textbf{141.71}  & 30.58 \\
        &         & L        &  & 5   & 1802.43        &  & 10          & 2700.46 & 16.31  &  & \textbf{13}  & \textbf{348.17}  & 16.31 \\ \cmidrule{2-14}
 &
  \multirow{3}{*}{5} &
  S &
   &
  3 &
  54.17 &
   &
  13 &
  151.62 &
  47.62 &
   &
  \textbf{15} &
  \textbf{44.61} &
  47.62 \\
        &         & M        &  & 5   & 193.67         &  & \textbf{15} & 587.69  & 45.53  &  & \textbf{15}  & \textbf{127.40}  & 45.53 \\
        &         & L        &  & 3   & 1093.78        &  & 11          & 2327.24 & 24.45  &  & \textbf{14}  & \textbf{505.07}  & 24.45 \\ \midrule
\multirow{6}{*}{50} &
  \multirow{3}{*}{2} &
  S &
   &
  6 &
  514.15 &
   &
  12 &
  1755.67 &
  37.98 &
   &
  \textbf{14} &
  \textbf{176.36} &
  37.98 \\
        &         & M        &  & 6   & 2064.98        &  & 4           & 3281.07 & 10.83  &  & \textbf{15}  & \textbf{273.79}  & 10.84 \\
        &         & L        &  & 1   & 3345.30        &  & 0           & 3600.00 & $-\infty$ &  & \textbf{15}  & \textbf{897.07}  & 70.61 \\ \cmidrule{2-14}
 &
  \multirow{3}{*}{5} &
  S &
   &
  4 &
  310.30 &
   &
  \textbf{14} &
  1370.53 &
  56.13 &
   &
  \textbf{14} &
  \textbf{117.69} &
  56.13 \\
        &         & M        &  & 13  & 1578.86        &  & 9           & 2707.62 & 0.00   &  & \textbf{14}  & \textbf{406.44}  & 0.00  \\
        &         & L        &  & 2   & 3254.23        &  & 1           & 3600.00 & $-\infty$ &  & \textbf{15}  & \textbf{1205.79} & 56.00 \\ \midrule
\multicolumn{3}{r}{Summary:} &  & 136 &                &  & 262         &         &        &  & \textbf{342} &                  &       \\ \bottomrule
\end{tabular}%
}
\caption{Comparison results for the NL model (joint estimation of $(\bbt,\bld)$).}
\label{tab:GeneralNL}
\end{table}

\subsubsection{Joint Estimating $(\bbt,\bld)$ for the TNL Model}
Table~\ref{tab:GeneralTreeNL} summarizes the performance of three optimization 
strategies on the TNL dataset across various problem sizes and tree structures. 
We consider the following methods:  

\begin{itemize}
    \item \textbf{\texttt{SLSQP}:} Applied directly to the estimation problem.  
    \item \textbf{\texttt{L-BFGS-B+SLSQP}:} Our two-step procedure in which the 
    outer step is solved by \texttt{SLSQP} (to handle the constraints on 
    $\blambda$ values between internal nodes and their children), while the 
    inner step is solved by \texttt{L-BFGS-B}.  
    \item \textbf{\texttt{ECP+SLSQP}:} Our proposed approach, combining \texttt{ECP} 
    in the inner step with \texttt{SLSQP} in the outer step.  
\end{itemize}

\begin{table}[htb]\footnotesize
\centering
\resizebox{\textwidth}{!}{%
\begin{tabular}{ccclrrlrrrlrrr}
\toprule
 &
   &
   &
   &
  \multicolumn{2}{c}{SLSQP} &
   &
  \multicolumn{3}{c}{L-BFGS-B+SLSQP} &
   &
  \multicolumn{3}{c}{ECP+SLSQP} \\ \cmidrule{5-6} \cmidrule{8-10} \cmidrule{12-14} 
\#Att &
  Tree &
  Size &
   &
  \#Best &
  AveTime(s) &
   &
  \#Best &
  AveTime(s) &
  AveGap(\%) &
   &
  \#Best &
  AveTime(s) &
  AveGap(\%) \\ \midrule
\multirow{6}{*}{5} &
  \multirow{3}{*}{2-2} &
  S &
   &
  13 &
  29.78 &
   &
  \textbf{15} &
  51.68 &
  0.01 &
   &
  \textbf{15} &
  \textbf{27.88} &
  0.01 \\
        &         & M        &  & \textbf{14} & \textbf{102.86} &  & 13          & 181.90  & 0.00    &  & 13           & 104.24           & 0.00  \\
        &         & L        &  & \textbf{15} & 502.16          &  & 14          & 614.19  & 0.00    &  & 14           & \textbf{313.84}  & 0.00  \\ \cmidrule{2-14}
 &
  \multirow{3}{*}{3-3} &
  S &
   &
  \textbf{14} &
  \textbf{42.19} &
   &
  13 &
  89.60 &
  0.00 &
   &
  13 &
  58.40 &
  0.00 \\
        &         & M        &  & 11          & \textbf{142.16} &  & \textbf{15} & 277.99  & 0.00    &  & \textbf{15}  & 237.07           & 0.00  \\
 &
   &
  L &
   &
  \textbf{14} &
  \textbf{971.60} &
   &
  \textbf{14} &
  1136.48 &
  0.00 &
   &
  \textbf{14} &
  1128.02 &
  0.00 \\ \midrule
\multirow{6}{*}{10} &
  \multirow{3}{*}{2-2} &
  S &
   &
  11 &
  54.15 &
   &
  \textbf{15} &
  102.32 &
  0.00 &
   &
  \textbf{15} &
  \textbf{39.58} &
  0.00 \\
        &         & M        &  & \textbf{13} & 234.50          &  & 12          & 507.94  & 0.00    &  & 12           & \textbf{221.72}  & 0.00  \\
        &         & L        &  & 13          & 1014.56         &  & 12          & 1380.87 & 0.00    &  & \textbf{14}  & \textbf{520.19}  & 0.00  \\ \cmidrule{2-14}
 &
  \multirow{3}{*}{3-3} &
  S &
   &
  7 &
  \textbf{70.54} &
   &
  \textbf{14} &
  155.01 &
  0.01 &
   &
  \textbf{14} &
  81.21 &
  0.01 \\
        &         & M        &  & 12          & 381.46          &  & \textbf{15} & 623.40  & 0.00    &  & \textbf{15}  & \textbf{379.52}  & 0.00  \\
        &         & L        &  & 12          & 1718.54         &  & 12          & 2406.47 & 0.00    &  & \textbf{14}  & \textbf{1487.45} & 0.00  \\ \midrule
\multirow{6}{*}{20} &
  \multirow{3}{*}{2-2} &
  S &
   &
  9 &
  148.32 &
   &
  14 &
  410.81 &
  0.00 &
   &
  \textbf{15} &
  \textbf{72.73} &
  0.00 \\
        &         & M        &  & 11          & 556.81          &  & \textbf{15} & 1378.57 & 0.00    &  & 13           & \textbf{321.47}  & 0.00  \\
        &         & L        &  & \textbf{14} & 2746.51         &  & 5           & 3304.86 & -0.01   &  & 13           & \textbf{1345.50} & 0.00  \\ \cmidrule{2-14}
 &
  \multirow{3}{*}{3-3} &
  S &
   &
  4 &
  120.84 &
   &
  \textbf{15} &
  360.24 &
  0.04 &
   &
  \textbf{15} &
  \textbf{115.13} &
  0.04 \\
        &         & M        &  & 9           & 695.52          &  & \textbf{15} & 1853.95 & 0.01    &  & 14           & \textbf{572.40}  & 0.00  \\
        &         & L        &  & 8           & 2823.47         &  & 5           & 3327.97 & -111.60 &  & \textbf{15}  & \textbf{2461.22} & 0.01  \\ \midrule
\multirow{6}{*}{50} &
  \multirow{3}{*}{2-2} &
  S &
   &
  7 &
  873.87 &
   &
  11 &
  2141.19 &
  0.00 &
   &
  \textbf{13} &
  \textbf{170.87} &
  0.00 \\
        &         & M        &  & 3           & 2498.82         &  & 2           & 3473.64 & 0.06    &  & \textbf{14}  & \textbf{1088.98} & 0.09  \\
        &         & L        &  & 0           & 3470.52         &  & 0           & 3600.00 & $-\infty$  &  & \textbf{15}  & \textbf{2563.97} & 35.89 \\ \cmidrule{2-14}
 &
  \multirow{3}{*}{3-3} &
  S &
   &
  6 &
  863.72 &
   &
  11 &
  2641.55 &
  $+\infty$ &
   &
  \textbf{15} &
  \textbf{337.35} &
  $+\infty$ \\
        &         & M        &  & 2           & 2383.00         &  & 1           & 3600.00 & 0.15    &  & \textbf{15}  & \textbf{1889.40} & 0.24  \\
        &         & L        &  & 0           & 2633.40         &  & 0           & 3600.00 & $-\infty$  &  & \textbf{15}  & \textbf{3339.22} & 24.13 \\ \midrule
\multicolumn{3}{r}{Summary:} &  & 222         &                 &  & 258         &         &         &  & \textbf{340} &                  &       \\ \bottomrule
\end{tabular}%
}
\caption{Comparison results for the NL model (joint estimation of $(\bbt,\bld)$).}
\label{tab:GeneralTreeNL}
\end{table}

The results demonstrate that \texttt{ECP+SLSQP} clearly outperforms the other methods, 
achieving the highest number of best solutions ($340/360$), with the largest average 
gaps and competitive runtimes. While \texttt{SLSQP} alone is efficient on small 
problems, it fails to maintain solution quality as the problem size increases. 
\texttt{L-BFGS-B+SLSQP} improves accuracy but often fails to converge or exceeds 
time limits on large instances, leading to infeasible or divergent solutions 
(e.g., infinite or negative gaps). In contrast, \texttt{ECP+SLSQP} remains stable 
and efficient even in the most challenging settings, demonstrating superior 
scalability, robustness, and solution quality.

\begin{figure}[htb]
    \centering
    \includegraphics[width=\linewidth]{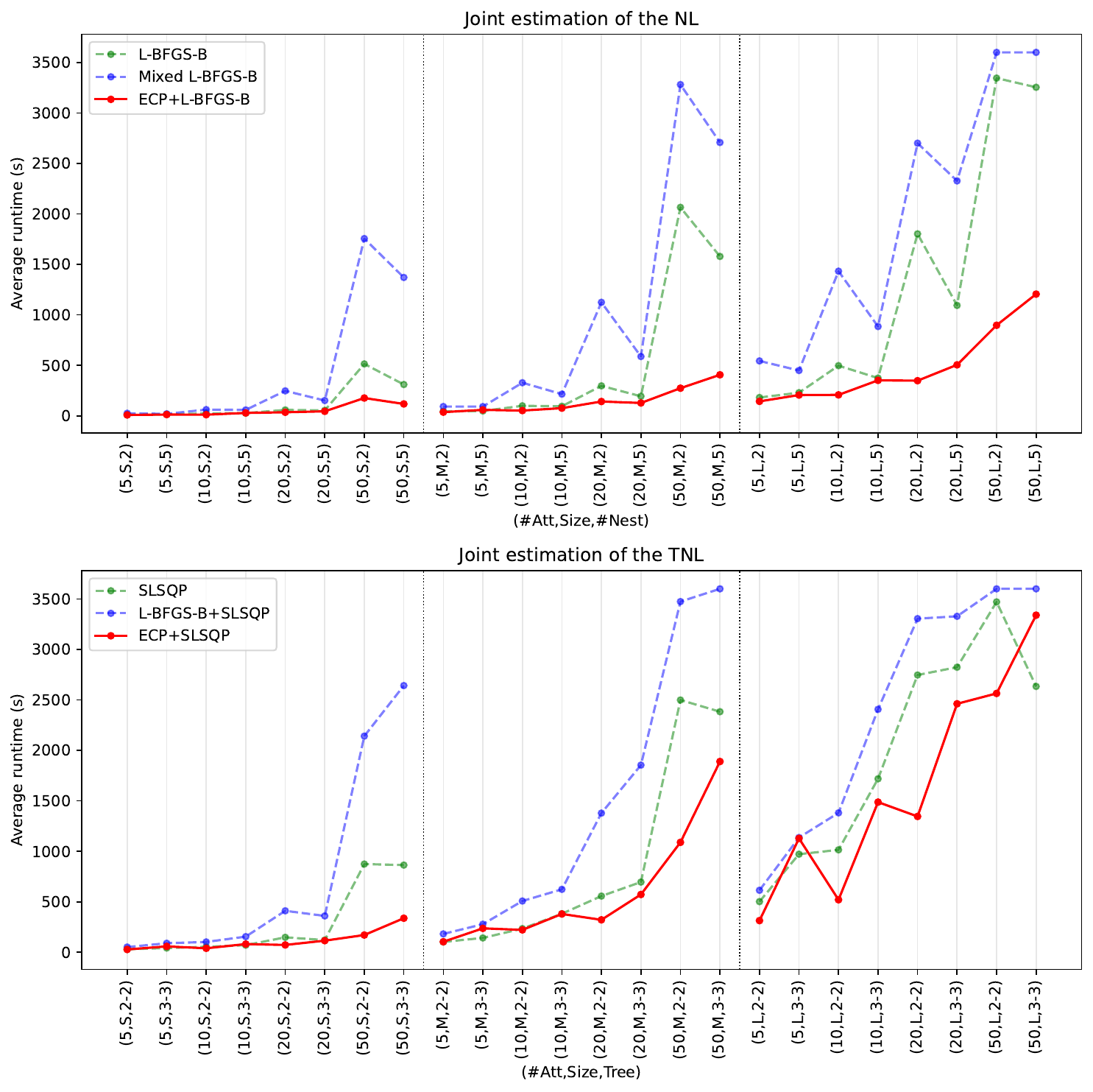}
    \caption{Comparison of average solving time across different methods for the joint estimation of the  NL and  TNL models.}
    \label{fig:avetime_size_general}
\end{figure}

Figure~\ref{fig:avetime_size_general} presents the results for joint estimation of the 
NL and TNL models, comparing the \texttt{SciPy} solvers with the exponential cone 
reformulation approaches. In both cases, ECP-based methods achieve significantly 
lower runtimes and exhibit superior scalability, clearly outperforming the 
\texttt{SciPy} solvers as model complexity increases.

Our experiments comprehensively evaluate the proposed exponential cone reformulation 
 approach against standard gradient-based solvers from \texttt{SciPy} across the 
MNL, NL, and TNL models. The results consistently demonstrate the superiority of ECP in 
terms of solution quality, reliability, and scalability. For the MNL model, ECP achieves 
a perfect optimization rate across all instances, while \texttt{L-BFGS-B} (best gradient-based method for the problem settings) fails on a 
subset of cases. The performance gap becomes even more pronounced for the more complex 
NL and TNL models: ECP-based methods reliably solve all instances, whereas the gradient-based 
solvers frequently fail to converge, return infeasible solutions, or hit time limits. 
In terms of runtime, ECP is not only significantly faster on medium and large-scale 
instances, but also scales more gracefully with increasing numbers of attributes, nests, 
and dataset sizes. Furthermore, for the joint estimation problems, our approaches such as \texttt{ECP+L-BFGS-B} and 
\texttt{ECP+SLSQP} leverage the strengths of the ECP solver, offering the best balance 
between accuracy and efficiency in joint estimation problems. Overall, the results 
highlight that ECP-based reformulations provide a robust and scalable framework for 
large-scale logit model estimation, outperforming state-of-the-art gradient-based solvers.

\section{Conclusion}  
\label{sec:concl}
We revisited parameter estimation for MNL, NL, and TNL models through convex conic 
optimization, showing that their maximum-likelihood problems can be reformulated as 
ECPs. Building on this insight, we analyze the polynomial-time computational complexity 
of solving the equivalent ECP reformulations for the MNL model, as well as for the 
NL and TNL models with fixed scale parameters.
we proposed a two-stage 
procedure that alternates between updating scale parameters and solving ECPs for 
utility coefficients.  Experiments on synthetic datasets demonstrate that ECP-based methods 
consistently outperform gradient-based solvers in terms of likelihood values, 
robustness, and runtime, especially on large-scale and high-dimensional instances.  
These results establish exponential-cone programming as a practical, scalable, and 
reliable alternative for discrete-choice model estimation.  

A natural extension is the development of conic optimization techniques for the 
\emph{global} estimation of nested logit,  cross-nested logit model, or generalized network-based choice models \citep{Train2009,Bier06, Mai16}. Unlike the 
fixed-parameter setting studied here, these formulations involve jointly optimizing 
over both utility coefficients and dissimilarity parameters, leading to highly 
non-convex landscapes. Additional challenges include ensuring identifiability of 
parameters, handling cross-nesting structures where alternatives belong to multiple 
nests, and designing scalable algorithms that can cope with the resulting 
high-dimensional optimization problems. Addressing these challenges would mark 
a significant step toward robust and efficient estimation of the most general 
forms of discrete-choice models.

\bibliographystyle{plainnat_custom}
\bibliography{refs}

\pagebreak
\appendix
\begin{center}
    {\Huge Appendix}
\end{center}

In this appendix, we provide the proofs omitted from the main text (Section~\ref{apd:proofs}) 
as well as detailed numerical analyses that support the experimental results reported in 
the main paper (Section \ref{apd:numerical-analyses}).

\section{Proofs}\label{apd:proofs}

\subsection{Proof of Proposition \ref{prop:ipm-complexity-mnl-short}}
\begin{proof} The complexity is obtained by analyzing both the number of Newton steps required in the path-following interior-point method and the computational cost incurred at each iteration.  

\paragraph{Bound on the number of interior-point iterations.}
The conic reformulation can be expressed in standard form $\max\{c^\top x : Ax=b,\; x\in\mathcal{K}\}$, where $\mathcal{K}$ is the product of 
(i) $Z$ exponential cones $\cK_{\exp}$, one for each pair $(n,j)$ with $j\in S_n$, and 
(ii) a small number of linear cones corresponding to slack variables from the constraints $\sum_{j\in S_n} z_{nj}\leq 1$. 
Since the exponential cone admits a self-concordant barrier with constant parameter $\nu_{\exp}=\Theta(1)$ \citep{Chares2009,BenTalNemirovski2001}, the total barrier parameter of the product cone is
\[
\nu = Z\times\,\nu_{\exp} + \nu_{\text{lin}} = \Theta(Z),
\]
where $\nu_{\text{lin}}$ accounts for the linear cones and is negligible compared to $Z\times\nu_{\exp}$. 
By the general theory of self-concordant barriers, a short-step or path-following primal--dual interior-point method therefore requires 
\[
\mathcal{O}\!\left(\sqrt{\nu}\,\log(1/\varepsilon)\right) 
= \mathcal{O}\!\left(\sqrt{Z}\,\log(1/\varepsilon)\right)
\]
Newton steps to compute an $\varepsilon$-optimal primal--dual solution \citep{NesterovNemirovskii1994,BenTalNemirovski2001}.

\paragraph{Per-step cost.}
At each Newton step the KKT system associated with the product cone separates into $Z$ local $3\times 3$ exponential-cone blocks coupled only through the global variables $(\bbt,\bt)$ and the $N$ linear constraints $\sum_{j\in S_n} z_{nj}\le 1$.
Eliminating the $Z$ local cone variables by block Gaussian elimination yields a Schur complement in the $(\bbt,\bt)$ variables of size $(p+N)\times(p+N)$.
Each pair $(n,j)$ contributes a rank-one (or small-rank) term proportional to $\ba_{nj}\ba_{nj}^\top$ into the $\bbt$--$\bbt$ block and simple couplings with $t_n$; assembling these contributions over all $(n,j)$ costs
\[
\sum_{n=1}^N \sum_{j\in S_n} \mathcal{O}(p^2) \;=\; \mathcal{O}(Z\,p^2).
\]
Factoring the dense Schur complement then costs $\mathcal{O}((p+N)^3)$ in the worst case.
Local cone updates are $\mathcal{O}(1)$ per cone (hence $\mathcal{O}(Z)$ in total) and are dominated when $p$ or $N$ is moderate-to-large.
Thus the per-iteration arithmetic cost is $\mathcal{O}(Z\,p^2+(p+N)^3)$.

\paragraph{Total complexity.}
Multiplying the per-step cost by the iteration bound $\mathcal{O}(\sqrt{Z}\log(1/\varepsilon))$ gives the stated overall complexity.
\end{proof}

\subsection{Proof of Proposition \ref{prop:ipm-nl-ecp}}
\begin{proof} 
Following \citet{Nesterov1994interior,BenTalNemirovski2001}, the overall complexity of solving \eqref{prob:NL-ECP} using a path-following interior-point method can be decomposed into two components: the number of Newton steps required and the computational cost of each iteration. 

For the path-following method, we first express \eqref{prob:NL-ECP} in conic standard form
$\max \{c^\top x : Ax=b,\; x \in \mathcal{K}\},$
where $\mathcal{K}$ is the direct product of the following cones:  
(i) one exponential cone $\cK_{\exp}$ for each $k_{njl}$ (a total of $Z$ such cones),  
(ii) one exponential cone $\cK_{\exp}$ for each $h_{nl}$ (a total of $\Lambda$ such cones), and  
(iii) linear cones corresponding to the $\Lambda+N$ linear inequalities 
$\sum_{j \in \cN_l \cap S_n} k_{njl} \leq 1$ and $\sum_{l \in L_n} h_{nl} \leq 1$.  

The exponential cone admits a self-concordant barrier with constant parameter $\nu_{\exp} = \Theta(1)$ \citep{Chares2009,BenTalNemirovski2001}. 
Since barrier parameters are additive under direct products, the total barrier parameter for $\mathcal{K}$ is
\[
\nu \;=\; (Z+\Lambda)\,\nu_{\exp} \;+\; \nu_{\text{lin}}
\;=\; \Theta(Z+\Lambda),
\]
where $\nu_{\text{lin}}$ is the number of  linear cones. 
As $\nu_{\text{lin}}$ scales only linearly with the number of small linear-cone blocks, it is dominated by the term $(Z+\Lambda)\nu_{\exp}$.

Thus, by the general self-concordant barrier theory for primal--dual path-following methods,
the number of Newton steps required to reach an $\varepsilon$-optimal solution is
$\mathcal{O}(\sqrt{\nu}\log(1/\varepsilon))$
\citep{NesterovNemirovskii1994,BenTalNemirovski2001}.
With $\nu=\Theta(Z+\Lambda)$, this yields
\(
\mathcal{O}\!\big(\sqrt{Z+\Lambda}\,\log(1/\varepsilon)\big)
\)
iterations.

We now  analyse the cost for each Newton step. The Newton step solves a KKT system with block structure induced by the product cone.
Eliminating the local $3$-dimensional exponential-cone variables
$\{k_{njl}\}$ and $\{h_{nl}\}$ via block Gaussian elimination produces a Schur complement in the
\emph{global} variables $(\bbt,\{z_{nl}\},\{y_n\})$ of size $(p+\Lambda+N)\times(p+\Lambda+N)$.
Each $(n,j)$ with $j\in S_n$ contributes a rank-one (or small-rank) update involving $\ba_{nj}\ba_{nj}^\top$
to the $\bbt$--$\bbt$ block and simple couplings with the corresponding $z_{nl}$;
assembling these contributions costs
\[
\sum_{n=1}^N\sum_{j\in S_n}\mathcal{O}(p^2) \;=\; \mathcal{O}(Z\,p^2).
\]
Factoring the dense Schur complement then costs $\mathcal{O}((p+\Lambda+N)^3)$ in the worst case.
Local cone updates are $\mathcal{O}(1)$ per cone (hence $\mathcal{O}(Z+\Lambda)$ overall) and are dominated when $p$ or $N$ is moderate to large.
Therefore, the per-iteration cost is
\(
\mathcal{O}(Z\,p^2 + (p+\Lambda+N)^3).
\)

Multiplying the per-iteration cost by the iteration bound gives the stated total arithmetic complexity $\mathcal{O}\!\Big(\sqrt{Z+\Lambda}\,\log(1/\varepsilon)\ \cdot\ \big(Z\,p^2 + (p+\Lambda+N)^3\big)\Big).$
\end{proof}

\subsection{Proof of Proposition \ref{prop:ipm-tnl-ecp}}

\begin{proof} 
Similar to the MNL and NL model estimation, following \citet{Nesterov1994interior,BenTalNemirovski2001}, the overall complexity of solving \eqref{prob:NL-ECP} using a path-following interior-point method can be decomposed into two components: the number of Newton steps required and the computational cost of each iteration. 

For the path-following method,  write \eqref{prob:TNL-ECP} in conic standard form $\max\{c^\top x: Ax=b,\ x\in\mathcal{K}\}$ where $\mathcal{K}$ is the product of:
(a) $E$ exponential cones $\cK_{\exp}$ (one per active edge $(k,s)$ and individual $n$),
(b) nonnegative orthants for the $\Gamma$ linear inequalities $1\ge\sum_{s}y^n_{ks}$,
together with affine equalities for the leaf bounds.
The exponential cone admits a self-concordant barrier with constant parameter $\nu_{\exp}=\Theta(1)$ \citep{Chares2009,BenTalNemirovski2001}.
Barrier parameters add under direct products; hence the total barrier parameter is
\[
\nu \;=\; E\times\,\nu_{\exp} \;+\; \nu_{\text{lin}} \;=\; \Theta(E),
\]
since the contribution $\nu_{\text{lin}}$ of the small linear cones is lower order relative to $E$. Thus,
by the general theory of self-concordant barriers for path-following interior-point,
the number of Newton steps to reach an $\varepsilon$-optimal solution is
$\mathcal{O}(\sqrt{\nu}\log(1/\varepsilon))$ \citep{NesterovNemirovskii1994,BenTalNemirovski2001}.
With $\nu=\Theta(E)$ this yields
$\mathcal{O}\!\big(\sqrt{E}\,\log(1/\varepsilon)\big)$ iterations.

\smallskip
For the cost of each Newton step,  we note that 
each step solves a KKT system with the block structure induced by the product cone.
Eliminating the local $3$-dimensional exponential-cone variables $\{y^n_{ks}\}$ by block Gaussian elimination produces a Schur complement in the \emph{global} variables $(\bbt,\{z_k^n\})$ of size $(p+\Gamma+Z)\times(p+\Gamma+Z)$.
Assembling the $\bbt$--$\bbt$ block requires
\[
\sum_{n=1}^N\sum_{k\in S_n}\mathcal{O}(p^2) \;=\; \mathcal{O}(Z\,p^2)
\]
operations because only the leaf constraints $z_k^n\ge \bbt^\top \ba_{nk}$ couple $\bbt$ to the rest of the system.
The remaining blocks involving the $z$-variables collect $\mathcal{O}(E)$ scalar contributions from the edge cones and are dominated when $p$ or $(\Gamma+Z)$ is moderate to large.
Factoring the dense Schur complement costs $\mathcal{O}((p+\Gamma+Z)^3)$ in the worst case.
Therefore the per-iteration cost is $\mathcal{O}(Z\,p^2 + (p+\Gamma+Z)^3)$.

Multiplying the per-iteration cost by the iteration bound gives the stated total arithmetic complexity $\mathcal{O}\!\Big(\sqrt{E}\,\log(1/\varepsilon)\ \cdot\ \big(Z\,p^2 + (p+\Gamma+Z)^3\big)\Big)\;$
as desired. 
\end{proof}

\section{Detailed Numerical Analyses}\label{apd:numerical-analyses}
\subsection{Comparison of Different Solvers in Scipy}
In this section, we evaluate the performance of different solvers available in the 
\texttt{SciPy} library on small-scale datasets with five attributes. For the NL and TNL 
models, we fix the number of nests to $2$ and adopt a $2$--$2$ tree structure, respectively.  

\begin{figure}[htb]
    \centering
    \begin{subfigure}[b]{\textwidth}
        \centering
        \includegraphics[width=\textwidth]{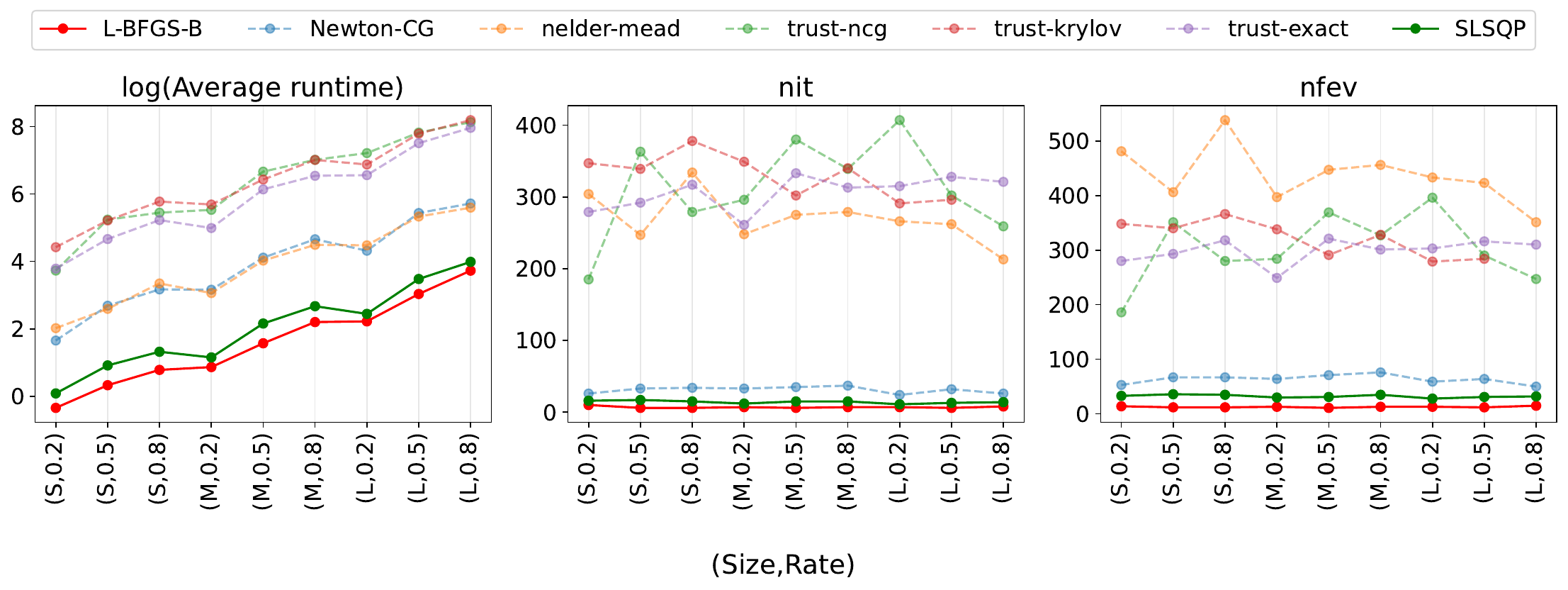}
        \caption{MNL instances}
        \label{fig:MNL}
    \end{subfigure}
    \begin{subfigure}[b]{\textwidth}
        \centering
        \includegraphics[width=\textwidth]{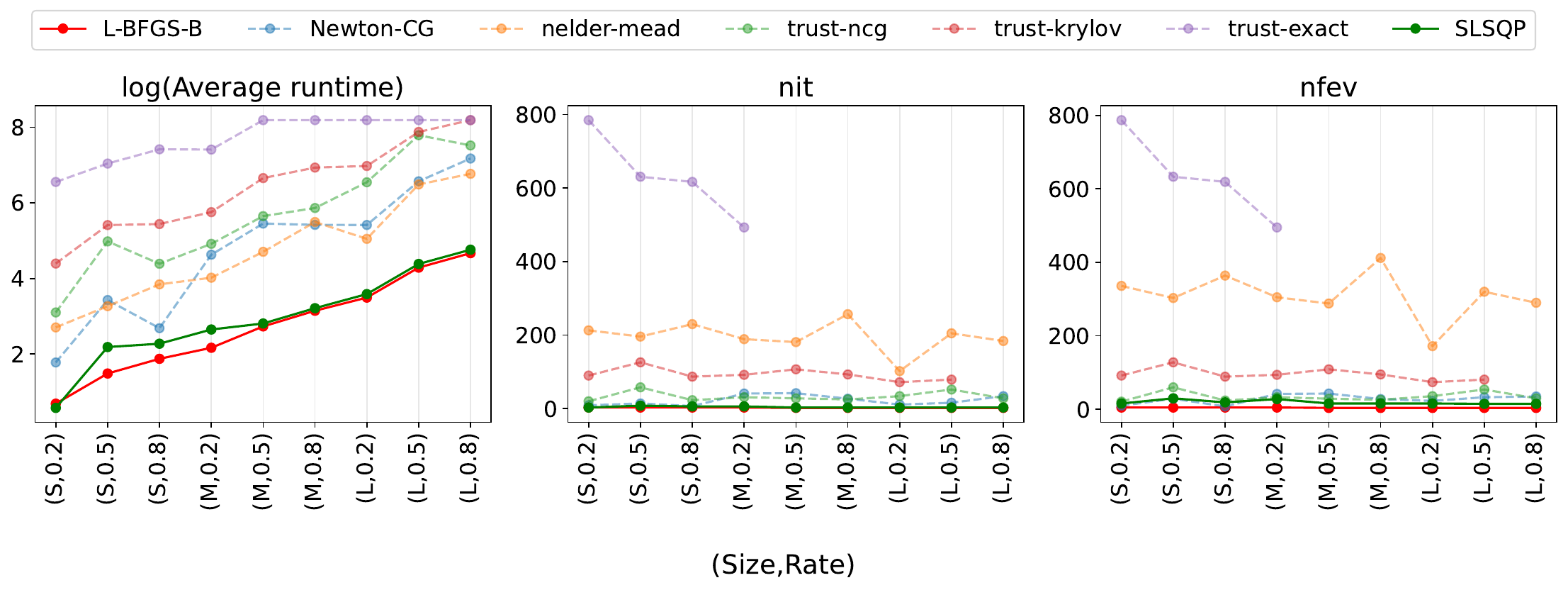}
        \caption{NL instances}
        \label{fig:NL}
    \end{subfigure}

    \begin{subfigure}[b]{\textwidth}
        \centering
        \includegraphics[width=\textwidth]{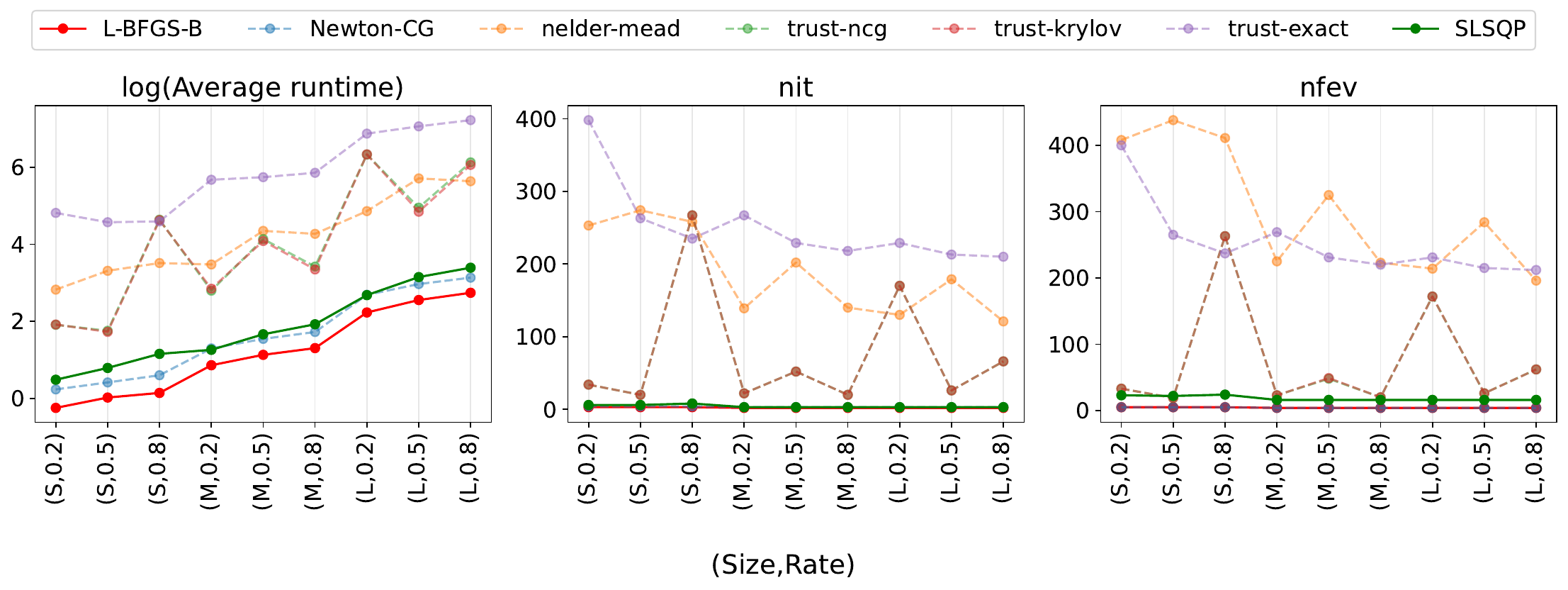}
        \caption{TNL instances}
        \label{fig:TNL}
    \end{subfigure}
    
    \caption{Performances of several common solvers in Scipy.}
    \label{fig:scipy_solver}
\end{figure}

Figures~\ref{fig:MNL} and \ref{fig:NL} show that the running times of 
\texttt{L-BFGS-B} (red line) and \texttt{SLSQP} (green line) remain the most stable and 
lowest as the dataset size increases from $S$ to $L$. Furthermore, both the number of 
iterations (\texttt{nit}) and the number of function evaluations (\texttt{nfev}) remain 
stable across dataset sizes, indicating that \texttt{L-BFGS-B} and \texttt{SLSQP} achieve 
the fastest convergence and highest stability for the MNL and NL models.  
For the TNL model (Figure~\ref{fig:TNL}), \texttt{L-BFGS-B} achieves the fastest running 
time, followed by \texttt{Newton-CG} and \texttt{SLSQP}. The values of \texttt{nit} and 
\texttt{nfev} for \texttt{L-BFGS-B} and \texttt{Newton-CG} are identical and the lowest 
among all methods, with \texttt{SLSQP} ranking next.  

Overall, \texttt{L-BFGS-B} is the most effective solver across all three models when 
$\blambda$ is fixed. When $\blambda$ is treated as a variable subject to additional 
constraints, \texttt{SLSQP} is employed, as it can handle such constraints while 
maintaining strong performance.

\subsection{Comparison Results across Different Choice Set Sizes}\label{append:choice_set}
\begin{figure}[htb]
    \centering
    \begin{subfigure}[b]{\textwidth}
        \centering
        \includegraphics[width=0.8\textwidth]{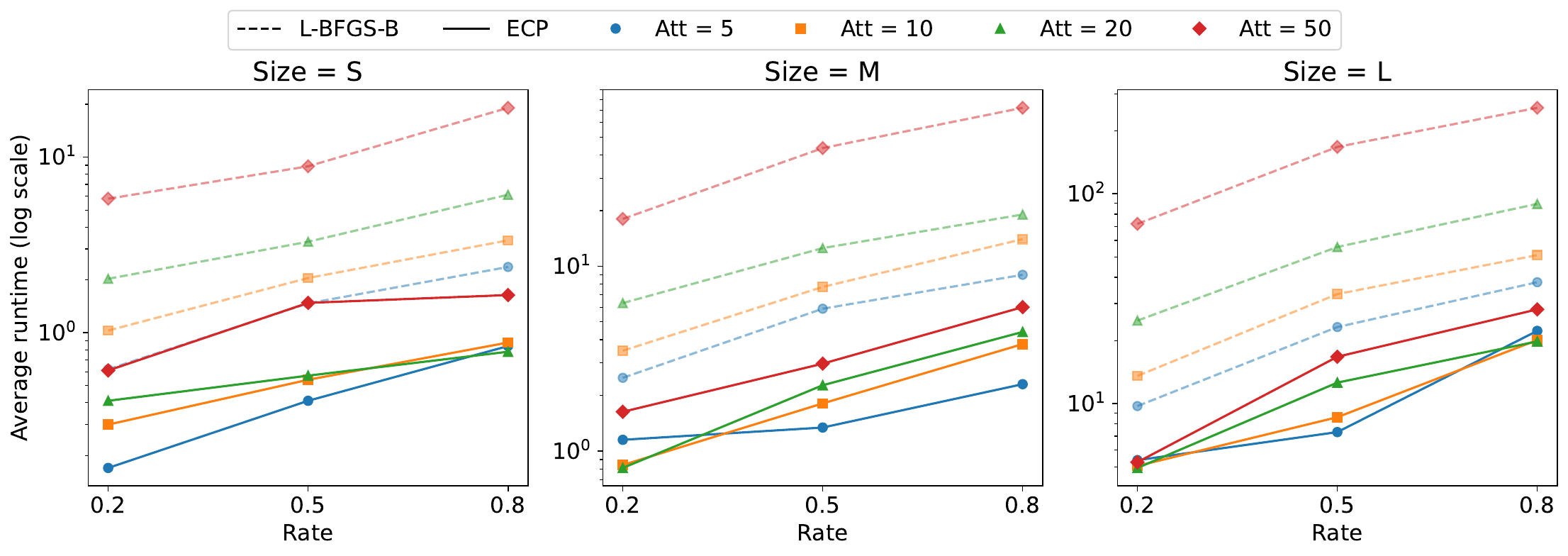}
        \caption{On the MNL instances}
        \label{fig:MNL_details}
    \end{subfigure}
    \begin{subfigure}[b]{\textwidth}
        \centering
        \includegraphics[width=0.8\textwidth]{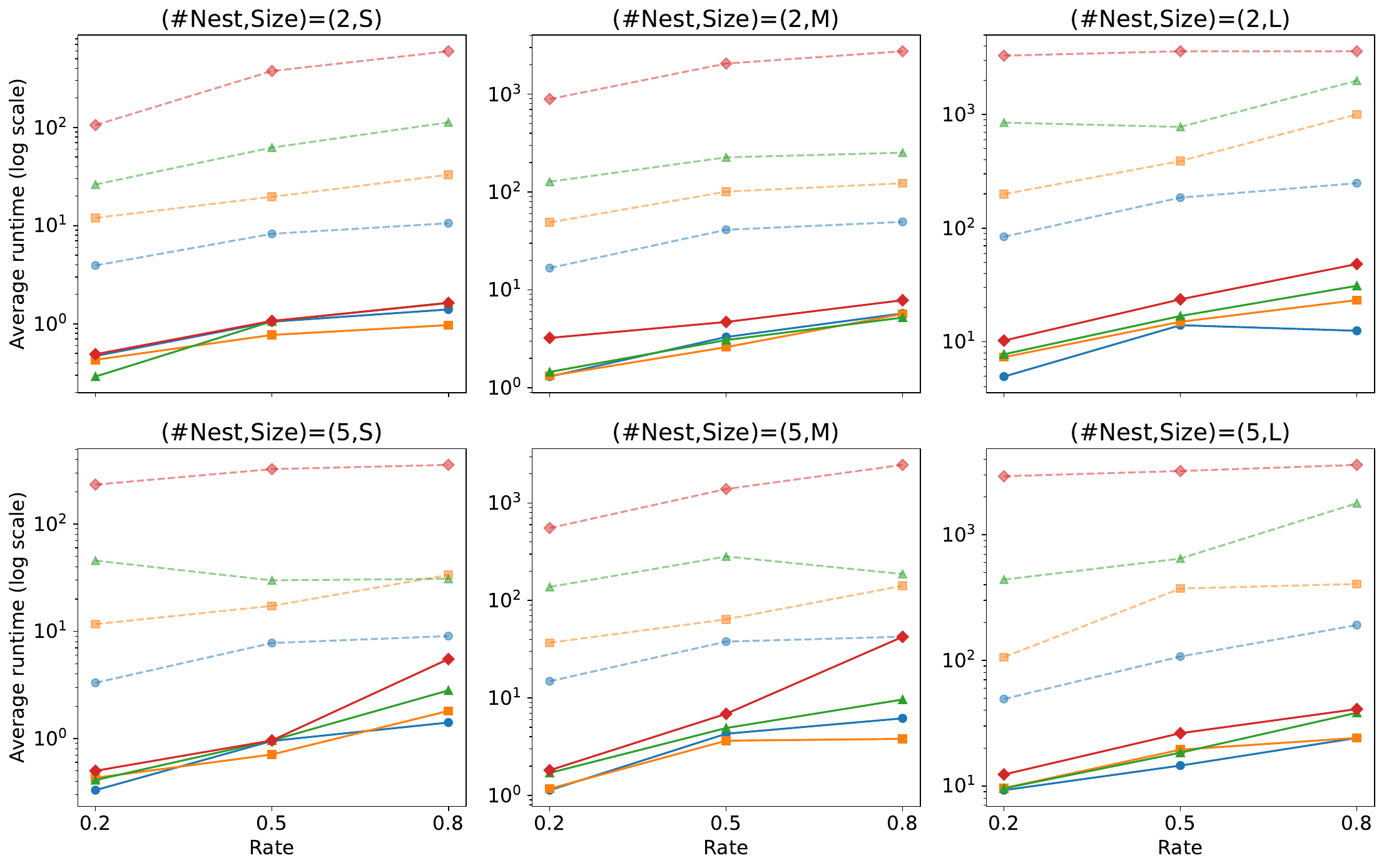}
        \caption{On the NL instances}
        \label{fig:NL_details}
    \end{subfigure}

    \begin{subfigure}[b]{\textwidth}
        \centering
        \includegraphics[width=0.8\textwidth]{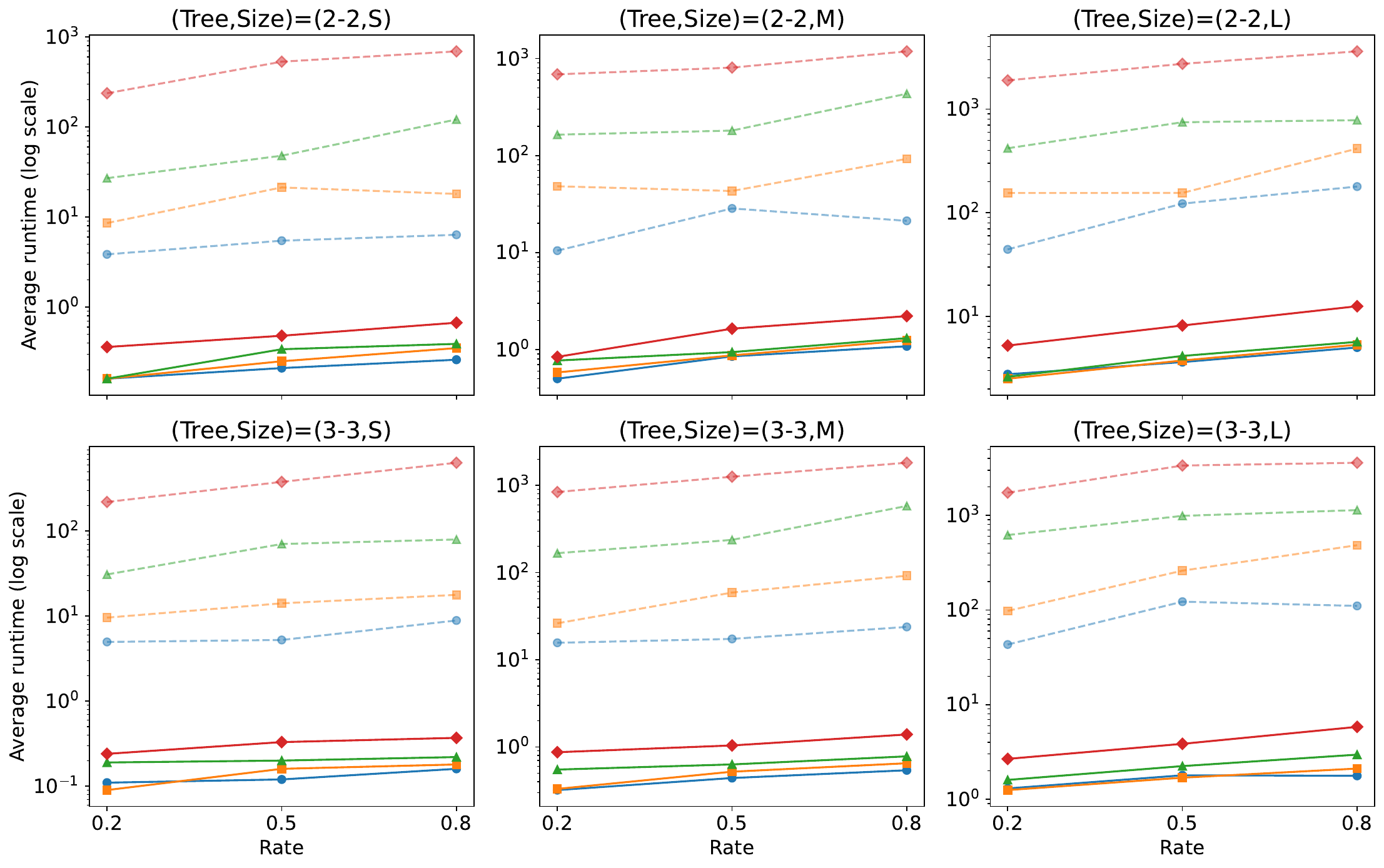}
        \caption{On the TNL instances}
        \label{fig:TNL_details}
    \end{subfigure}
    
    \caption{Performances of the ECP methods.}
    \label{fig:ECP_all}
\end{figure}

    Figures \ref{fig:ECP_all} visualizes the detailed results of Table \ref{tab:MNL_NL_TreeNL} when the datasets are grouped by choice set size $|S_n|$, where $|S_n|=$ \texttt{Rate}$\times m$ and \texttt{Rate} $\in \{0.2, 0.5, 0.8\}$. Here, we compare only the average solving time of \texttt{L-BFGS-B} and ECP over five instances per each set in 36 sets and do not report the number of optimally solved instances, as ECP solves all cases. Solving times for datasets with different numbers of attributes are represented by different colors: the runtime of \texttt{L-BFGS-B} is shown by dotted lines, while that of ECP is shown by solid lines of the same color. The results clearly indicate that, across all three datasets (MNL, NL, and TNL), ECP consistently outperforms \texttt{L-BFGS-B} in terms of runtime. In addition, the upward trend of the curves demonstrates that larger choice set sizes increase problem complexity and require longer solving times. Examining each method separately, the dotted lines of \texttt{L-BFGS-B} are distinctly separated as the number of attributes increases, whereas the lines of ECP are much closer together and occasionally overlap. This suggests that the computation time of \texttt{L-BFGS-B} is more sensitive to the number of attributes than that of ECP.



\begin{figure}[htb]
    \centering
    \begin{subfigure}[b]{\textwidth}
        \centering
        \includegraphics[width=0.75\textwidth]{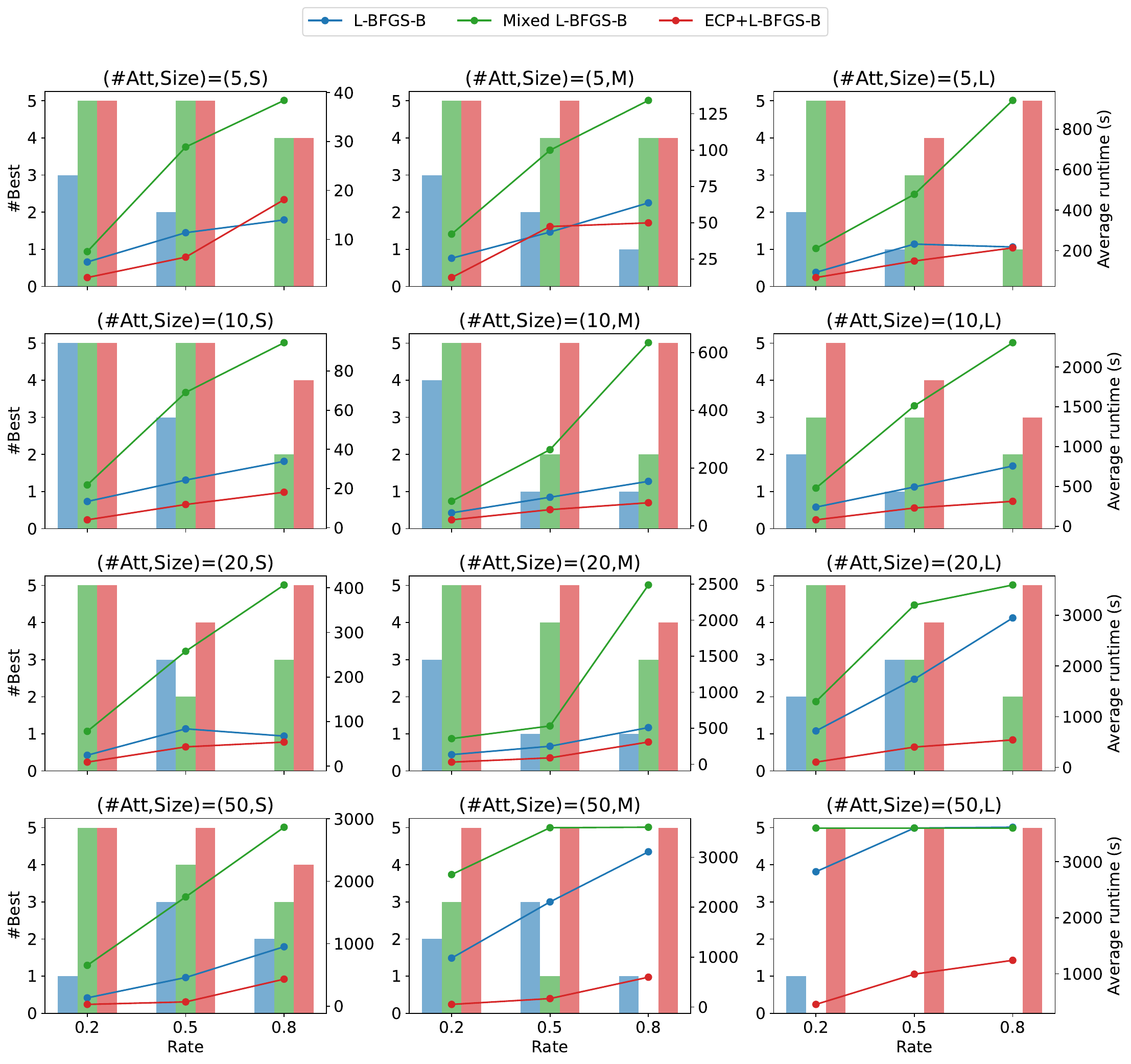}
        \caption{2 nests}
        \label{fig:GeneralNL_2}
    \end{subfigure}
    \begin{subfigure}[b]{\textwidth}
        \centering
        \includegraphics[width=0.75\textwidth]{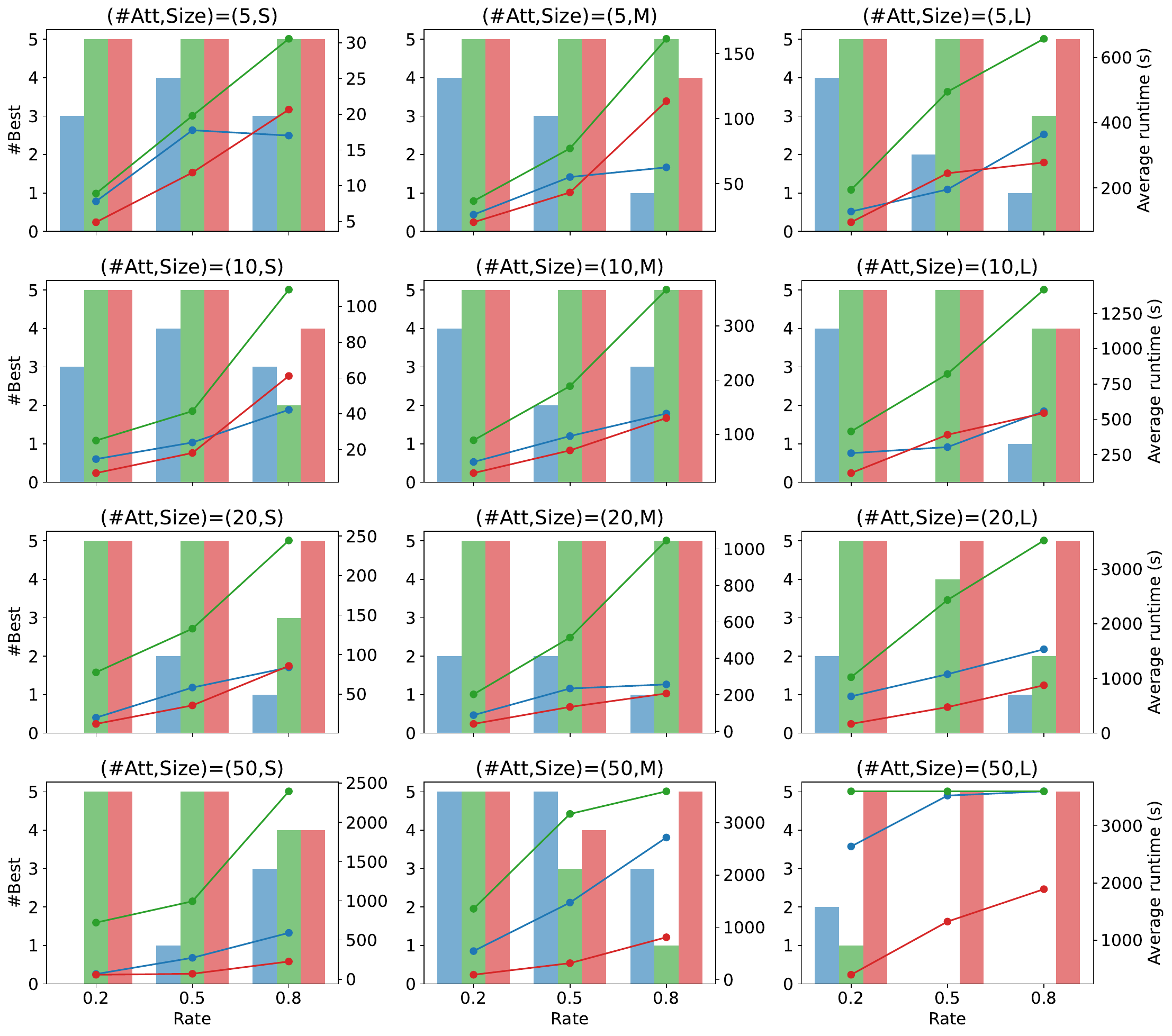}
        \caption{5 nests}
        \label{fig:GeneralNL_5}
    \end{subfigure}
    
    \caption{Performances of the ECP methods on the joint estimation of the NL datasets.}
    \label{fig:ECP_GeneralNL}
\end{figure}

\begin{figure}[htb]
    \centering
    \begin{subfigure}[b]{\textwidth}
        \centering
        \includegraphics[width=0.75\textwidth]{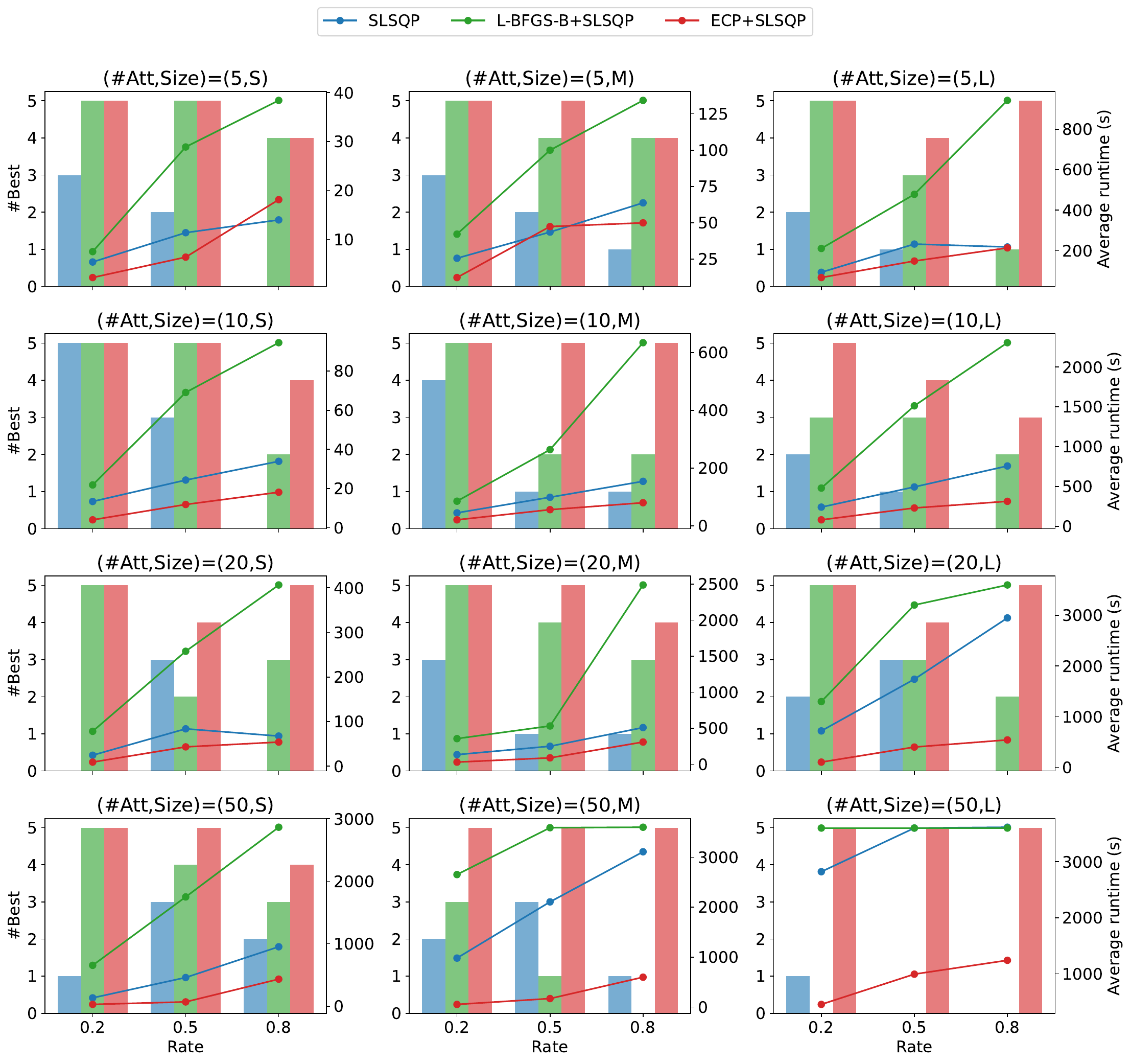}
        \caption{Tree 2-2}
        \label{fig:GeneralTNL_22}
    \end{subfigure}
    \begin{subfigure}[b]{\textwidth}
        \centering
        \includegraphics[width=0.75\textwidth]{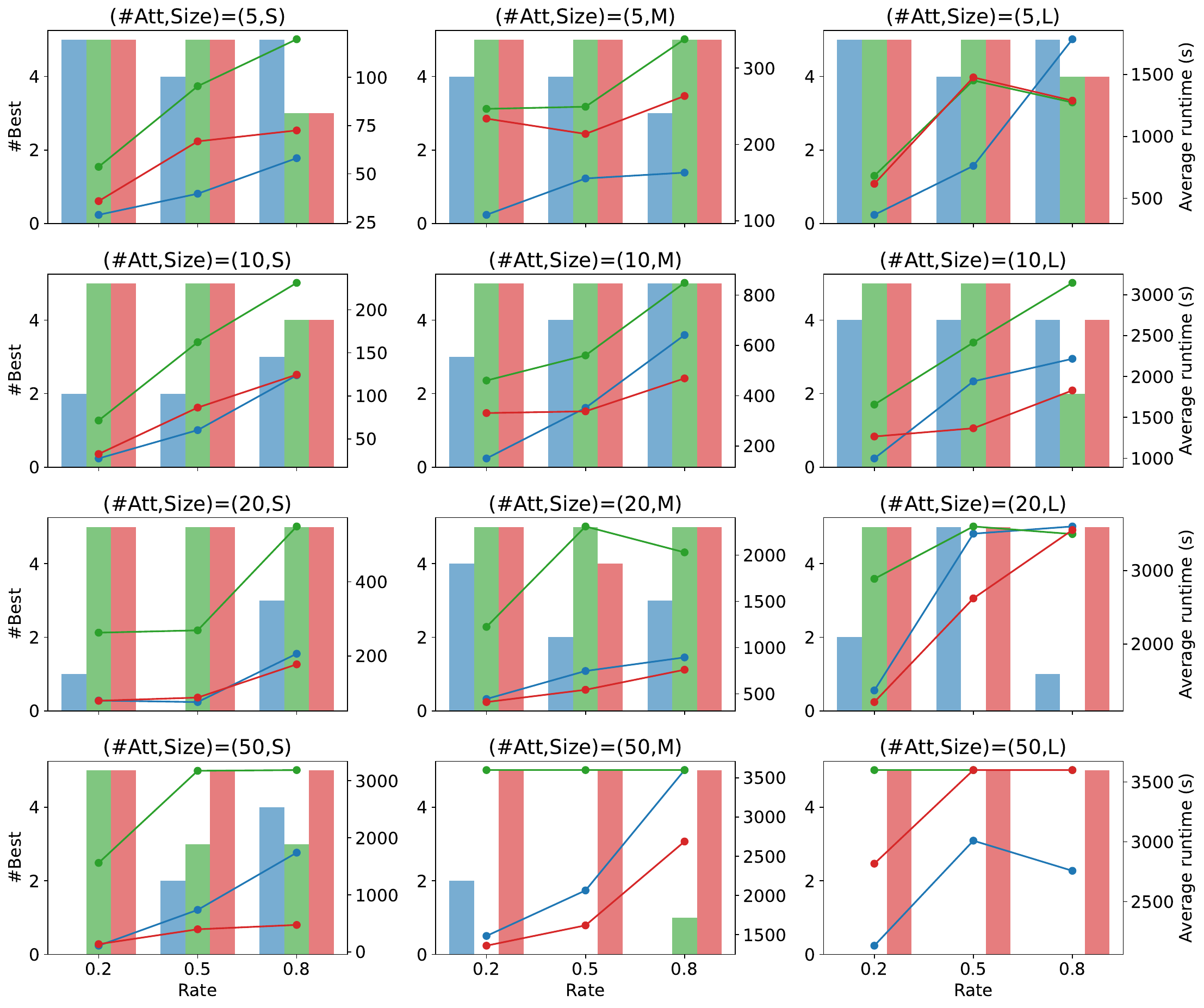}
        \caption{Tree 3-3}
        \label{fig:GeneralTNL_3-3}
    \end{subfigure}
    
    \caption{Performances of the ECP methods on the joint estimation of the TNL datasets.}
    \label{fig:ECP_GeneralTNL}
\end{figure}

Figures \ref{fig:ECP_GeneralNL} and \ref{fig:ECP_GeneralTNL} visualize the comparison results corresponding to Tables \ref{tab:GeneralNL} and \ref{tab:GeneralTreeNL}, respectively. These figures compare the number of best solutions obtained by each method as well as the average solving time across five instances per dataset. The number of best solutions provided by each method on each dataset is represented by a bar, while the average solving time is shown by the line of the same color. For both datasets involving the joint estimation of the NL and TNL models, using two-step procedures yield more best solutions than directly applying a single \texttt{SciPy} solver, with the combination of ECP and a \texttt{SciPy} solver being the most effective. However, the runtime of the two-stage procedure that employs two \texttt{SciPy} solvers is the highest in most cases and rapidly approaches the time limit as the \texttt{Rate} increases from 0.2 to 0.8. Methods that use ECP in the inner step demonstrate clear time advantages on the joint estimation of NL and 2-2 TNL datasets, and slightly outperform the single \texttt{SciPy} solver in the case of 3-3 TNL.



\end{document}